\documentclass[useAMS,usenatbib]{mn2e}

\usepackage{graphicx,amssymb,amstext} 

\begin{document}

\title[The AKARI All-Sky Survey at 90 micron]{Timeline analysis and wavelet multiscale analysis of the AKARI All-Sky Survey at 90 micron} 
\author[Lingyu Wang et al.]{
\parbox[t]{\textwidth}{
Lingyu Wang$^{1}$\thanks{E-mail: lingyu.wang@imperial.ac.uk}, Michael Rowan-Robinson$^{1}$, Issei Yamamura$^{2}$, Hiroshi Shibai$^{3}$, Rich Savage$^{4}$, Seb Oliver$^{4}$, Matthew Thomson$^{4}$, Nurur Rahman$^{4}$, Dave Clements$^{1}$, Elysandra Figueredo$^{5}$, Tomotsugu Goto$^{2}$, Sunao Hasegawa$^{2}$, Woong-Seob Jeong$^{2,8}$, Shuji Matsuura$^{2}$,
Thomas G. M\"{u}ller$^{7}$, Takao Nakagawa$^{2}$, Chris P. Pearson$^{6,9}$, Stephen Serjeant$^{5}$,
Mai Shirahata$^{2}$, Glenn J. White$^{5,6}$}
\\
$^{1}$Astrophysics Group, Blackett Laboratory, Imperial College of Science Technology and Medicine, London SW7 2BZ, UK\\
$^{2}$Institute of Space and Astronautical Science, Japan Aerospace Exploration Agency, Yoshinodai 3-1-1, Sagamihara, Kanagawa 229-8510, Japan\\
$^{3}$Graduate School of Sciences, Nagoya University, Furo-cho, Chikusa-ku, Nagoya 464-8602, Japan\\ 
$^{4}$Astronomy Centre, Department of Physics and Astronomy, Univeristy of Sussex, UK\\
$^{5}$Department of Physics and Astronomy, The Open University, Milton Keynes MK7 6AA, UK\\
$^{6}$Space Science \& Technology Department, CCLRC Rutherford Appleton Laboratory, Chilton, Didcot, Oxfordshire OX11 0QX, UK\\
$^{7}$Max Planck Institute for Extraterrestrial Physics, Giessenbachstrasse, 85748 Garching, Germany\\
$^{8}$Space Science Division, Korea Astronomy \& Space Science Institute (KASI), 61-1, Whaam-dong, Yuseong-gu, Deajeon, 305-348, South Korea\\
$^{9}$Department of Physics, University of Lethbridge, 4401 University Drive, Lethbridge, Alberta T1J 1B1, Canada
}

\date{}

\maketitle

\begin{abstract}
We present a careful analysis of the point source detection limit of the AKARI All-Sky Survey in the WIDE-S 90\,$\mu$m band near the North Ecliptic Pole (NEP). Timeline Analysis is used to detect IRAS sources and then a conversion factor is derived to transform the peak timeline signal to the interpolated 90\,$\mu$m flux of a source. Combined with a robust noise measurement, the point source flux detection limit at S/N $>5$ for a single detector row is $1.1\pm0.1$ Jy which corresponds to a point source detection limit of the survey of $\sim$0.4 Jy. 

Wavelet transform offers a multiscale representation of the Time Series Data (TSD). We calculate the continuous wavelet transform of the TSD and then search for significant wavelet coefficients considered as potential source detections. To discriminate real sources from spurious or moving objects, only sources with confirmation are selected. In our multiscale analysis, IRAS sources selected above $4\sigma$ can be identified as the only real sources at the Point Source Scales. We also investigate the correlation between the non-IRAS sources detected in Timeline Analysis and cirrus emission using wavelet transform and contour plots of wavelet power spectrum. It is shown that the non-IRAS sources are most likely to be caused by excessive noise over a large range of spatial scales rather than real extended structures such as cirrus clouds. 
\end{abstract}


\section{INTRODUCTION}
AKARI (previously known as ASTRO-F) is a Japanese infrared astronomical mission launched in February 2006 (Murakami et al. 2007). Its primary goal is to survey the entire sky with wide spectral coverage and high spatial resolution. It has two focal plane instruments, the Far-Infrared Surveyor (FIS, Kawada et al. 2007) covering a wavelength region from 50 - 180\,$\mu$m, and the Infrared Camera (IRC, Onaka et al. 2007) covering a wavelength region from 1.8 - 26.5\,$\mu$m. The FIS observes the sky in four photometric bands, referred to as N60, WIDE-S, WIDE-L and N160. In Table~\ref{FIS}, we give band name, band centre, detector array format and detector pixel size for each band. In comparison, IRAS has a pixel size of 1.5$\times$4.7 arcmin$^2$ at 60\,$\mu$m and 3.0$\times$5.0 arcmin$^2$ at 100\,$\mu$m.

\begin{table*}
\caption{Specifications of the Far-Infrared Surveyor (FIS).}\label{FIS}
\begin{tabular}[pos]{llllll}
BAND         & N60   & WIDE-S      & WIDE-L  & N160  & \\
\hline
band centre  & 65    &90           &140      &160    &[$\mu$m]\\
detector array format & 20$\times$2 & 20$\times$3 &15$\times$3 &15$\times$2&[row$\times$column]\\
detector pixel size & 26.8$\times$26.8 &26.8$\times$26.8 &44.2$\times$44.2  &44.2$\times$44.2& [arcsec$^2$]\\
\end{tabular}
\end{table*}

The first phase of the AKARI All-Sky Survey started in May 2006, lasting for one half year. About $70\%$ of the sky was observed by at least two independent scans during this period. In the subsequent complementary phase which ended on August 26th, 2007, due to exhaustion of helium, the sky coverage was increased to about $94\%$. Similarly, IRAS surveyed $98\%$ of the sky with a single scan, $96\%$ with two confirming scans and $72\%$ with three or more scans (Moshir et al. 1992).

The pre-flight $5\sigma$ point source detection limit for a single scan (with 2 or 3 detector rows) was estimated to be 0.6, 0.2, 0.4 and 0.8 Jy in the N60, WIDE-S, WIDE-L and N160 band respectively. The current in-orbit estimate based on noise measurement and absolute calibration using well-modelled objects is 2.4, 0.55, 1.4 and 6.3 Jy, with uncertainties $\leq30\%$ (Kawada et al. 2007). Thus, there is a factor of 3 or more degradation in all bands caused by excessive in-orbit noise, frequent glitches etc. On the other hand, the IRAS Point Source Catalogue (PSC) is complete to about 0.6 Jy at 60\,$\mu$m and about 1.0 Jy at 100\,$\mu$m in unconfused regions. At a $90\%$ completeness, the IRAS Faint Source Catalogue (FSC) achieved a depth of $\sim$0.2 Jy at 60\,$\mu$m by coadding the data after point-source filtering. The IRAS 100\,$\mu$m band to which the AKARI WIDE-S 90\,$\mu$m band is closer was not considered for catalogue qualification because of severe cirrus contamination. For the same reason, all IRAS flux densities at 100\,$\mu$m were declared to be moderate quality. The FSC usually contains sources with 100\,$\mu$m flux densities above 1 - 2 Jy. 

In this paper, we focus on confirmed detections from the timelines\footnote{The one-dimensional data stream in a single detector pixel is defined as a timeline.} to give a reliable and updated point source detection limit estimate in the WIDE-S 90\,$\mu$m band, taking advantage of the improved data reduction pipeline. Section~\ref{data} describes the AKARI Performance Verification (PV) phase data near the North Ecliptic Pole (NEP), robust noise measurement and 90\,$\mu$m flux interpolation. Section~\ref{timeline} describes the Timeline Analysis processing, the conversion factor and the association between the seconds-confirmed (SCON) and hours-confirmed (HCON)\footnote{Following the nomenclature of IRAS, seconds-confirmed sources are detected in at least two detector rows in a single scan; Hours-confirmed sources are detected in at least two independent scans which are hours apart from each other. Confirmation is essential to detect true fixed sources.} sources which are not found in the IRAS catalogues and infrared cirrus emission. A wavelet multiscale analysis of the selected IRAS sources is presented in Section~\ref{wavelets}, which also includes an investigation on the extended nature of the non-IRAS sources. Finally, we give our conclusions in Section~\ref{conclusions}.

\begin{figure}
\includegraphics[height=2.2in,width=3.2in]{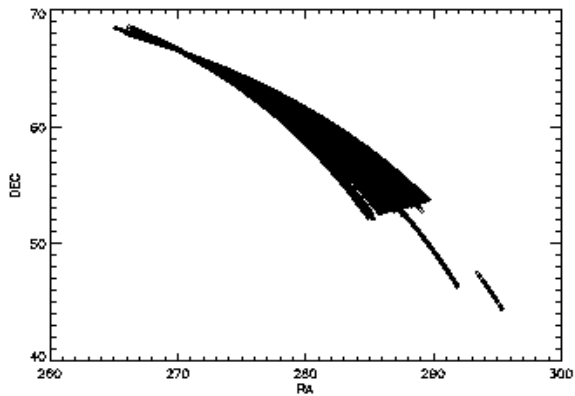}
\caption{The area covered by the AKARI PV phase data near the NEP.}
\label{fig:coverage}
\end{figure}

\section{THE DATA}
\label{data}
We have used 412 pieces of scan data near the NEP each of which is of one minute length, taken during the PV phase from April 24th 2006 to May 7th 2006. The data file is in a dedicated FITS format called Time Series Data (TSD) format. It has been processed and calibrated by the May 2007 version of the Green Box which is part of a dedicated data reduction pipeline for the AKARI All-Sky Survey (Yamamura et al. in prep.). The positional information is provided by the ground-based attitude determination system (GADS) and the rms positional accuracy is $\sim$30$''$. Eventually, the positional accuracy will be improved to a few arcseconds using ESA's pointing reconstruction processing. The data covers approximately 40 square degrees (Fig.~\ref{fig:coverage}) and $\sim$70 sources in the IRAS FSC with 60\,$\mu$m flux densities ranging from $\sim$0.2 to $10.1$ Jy, excluding the Cat's Eye Nebula. This group of objects forms an important base to estimate the sensitivity, reliability and completeness of the AKARI All-Sky Survey. 

To compare with previous absolute calibration of the AKARI All-Sky Survey data, we have used another data set containing 9 asteroids whose positions, observed epochs, and predicted monochromatic 90\,$\mu$m flux densities based on model spectral energy distributions (SEDs; M\"{u}eller \& Lagerros 2002; M\"{u}eller et al. 2008) are listed in Table~\ref{catAster}.

\begin{figure}
\includegraphics[height=2.2in,width=3.2in]{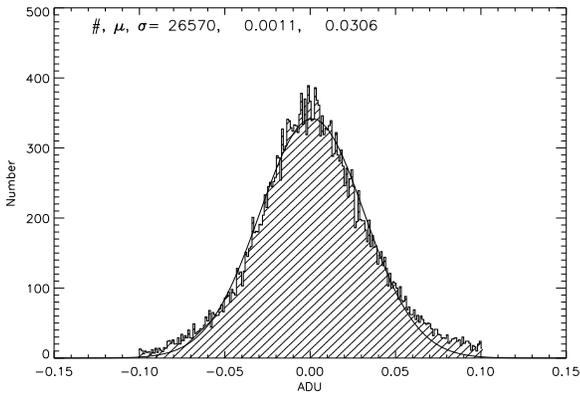}
\caption{The final data histogram for a typical TSD file. \# is the number of data points. $\mu$ and $\sigma$ is the mean and sigma of the fitted Gaussian function respectively.}
\label{fig:noise}
\end{figure}

\begin{figure}
\includegraphics[height=2.2in,width=3.2in]{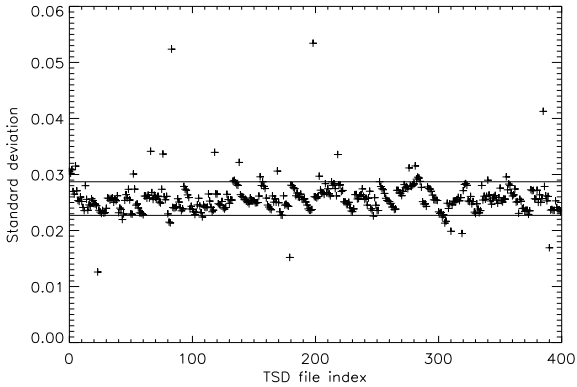}
\caption{The standard deviation distribution in the AKARI PV phase data near the NEP. The median value of $\sigma$ is 0.025 ADU.}
\label{fig:noisedist}
\end{figure}

\begin{table*}
\caption{Asteroids used in absolute calibration.}\label{catAster}
\begin{tabular}[pos]{rrrrrrr}
ID     & NAME           & RA (J2000)    & DEC (J2000)   & $F_{90}^{model}$ (Jy) & DATE  & TIME\\
\hline
0210   & 6 Hebe         & 309.970 & -8.744  & 13.197 &2006.04.30 &14:55:29\\
0220   & 511 Davida     & 315.824 & -19.530 & 11.373 &2006.05.03 &04:02:44\\
0240   & 1 Ceres        & 328.007 & -21.064 & 172.95 &2006.05.14 &00:51:00\\
0295   & 230 Athamantis & 202.658 & -11.580 & 3.419 &2006.07.18 &01:56:30\\
0480   & 511 Davida     & 307.679 & -28.328 & 12.997 &2006.10.26 &16:20:22\\
0510   & 88 Thisbe      & 133.914 & 15.669  & 7.271 &2006.11.04 &14:31:57\\
0530   & 6 Hebe         & 319.527 & -23.694 & 23.311 &2006.11.07 &06:09:23\\
0540   & 1 Ceres        & 322.469 & -26.277 & 160.33 &2006.11.09 &01:08:33\\
0560   & 93 Minerva     & 152.718 & 19.193  & 4.619 &2006.11.20 &09:13:18\\
\end{tabular}
\end{table*}

\subsection{Robust noise estimation}\label{noiseest}
We use a $3\sigma$ clipping method to estimate the Gaussian noise in the data. The main processing steps are:

(1) For each detector data stream, we calculate and subtract the median with a specified width of 192 discrete samplings which corresponds to $\sim$8 seconds or $\sim$28.8 arcmin (background subtraction).  

(2) We discard flagged/bad data which include signals caused by resetting, calibration lamp, dead pixels, cosmic-ray glitches, etc. In addition, we delete any region affected by the Cat's Eye Nebula which is a very bright extended source. 

(3) A 2-point boxcar smoothing is applied to the timeline in each detector pixel. The smoothing length is equivalent to an angular size of $\sim$15 arcseconds. 

(4) We estimate the standard deviation and then exclude any signal with S/N $>3$. 

(5) We repeat step (3) until the standard deviation converges. The final data histogram is fitted with a Gaussian curve to derive the mean $\mu$ and the standard deviation $\sigma$.

Fig. \ref{fig:noise} shows the final data histogram of a typical TSD file or scan and its Gaussian fit.  Averaged over 412 scans, $\langle\sigma\rangle$ is $0.026\pm0.003$ ADU. After excluding outliers defined as $|\sigma-\langle\sigma\rangle|>0.003$ ADU (plus signs outside the region delineated by the two solid lines in Fig.~\ref{fig:noisedist}), the mean standard deviation $\langle\sigma\rangle$ is reduced to $0.025\pm0.001$ ADU.

\begin{figure*}
\begin{minipage}{.1\textwidth}\centering
\includegraphics[width=1.0\textwidth]{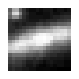}
\end{minipage}
\begin{minipage}{.1\textwidth}\centering
\includegraphics[width=1.0\textwidth]{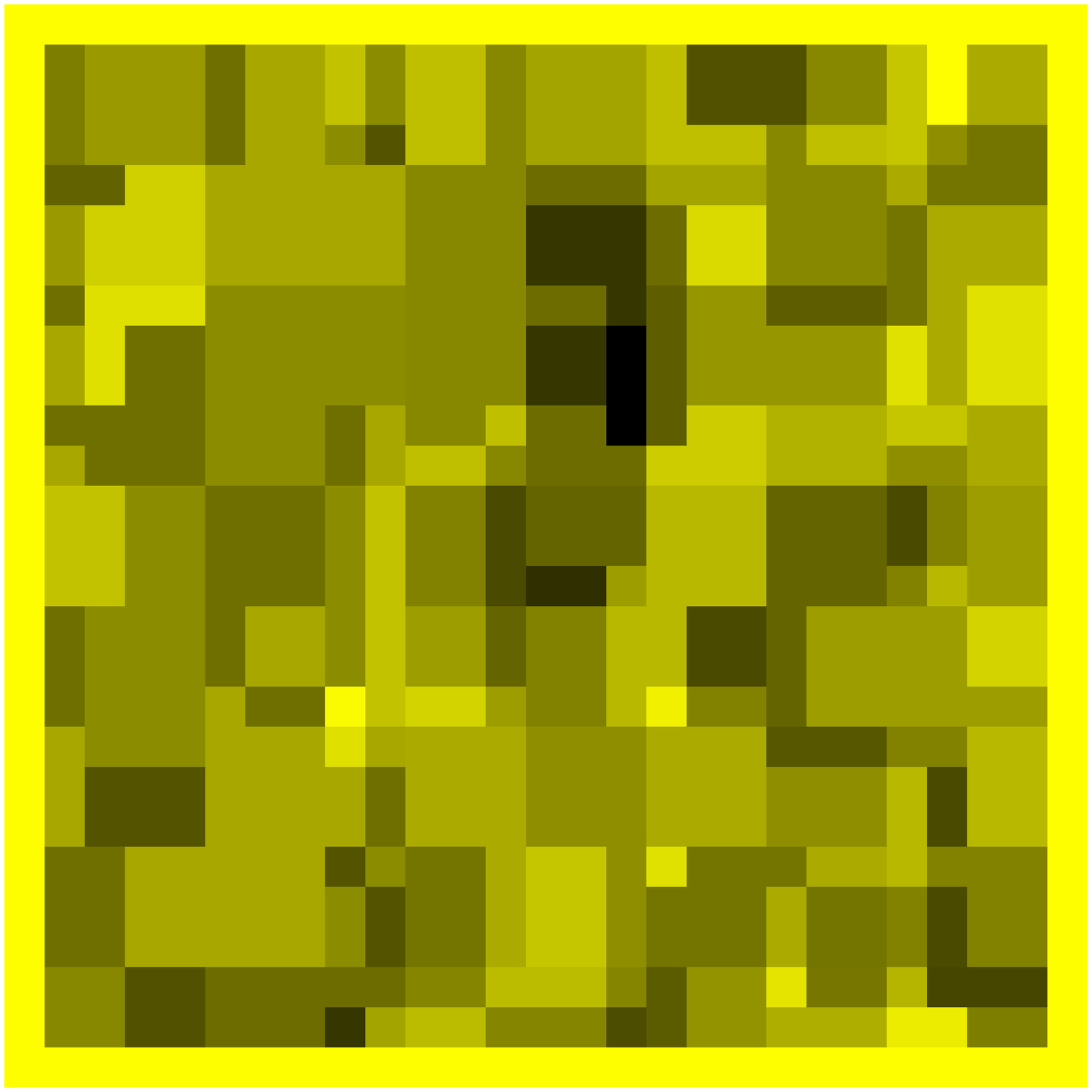}
\end{minipage}
\begin{minipage}{.1\textwidth}\centering
\includegraphics[width=1.0\textwidth]{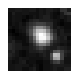}
\end{minipage}
\begin{minipage}{.1\textwidth}\centering
\includegraphics[width=1.0\textwidth]{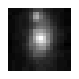}
\end{minipage}
\begin{minipage}{.1\textwidth}\centering
\includegraphics[width=1.0\textwidth]{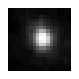}
\end{minipage}
\begin{minipage}{.1\textwidth}\centering
\includegraphics[width=1.0\textwidth]{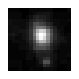}
\end{minipage}
\begin{minipage}{.1\textwidth}\centering
\includegraphics[width=1.0\textwidth]{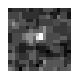}
\end{minipage}
\begin{minipage}{.1\textwidth}\centering
\includegraphics[width=1.0\textwidth]{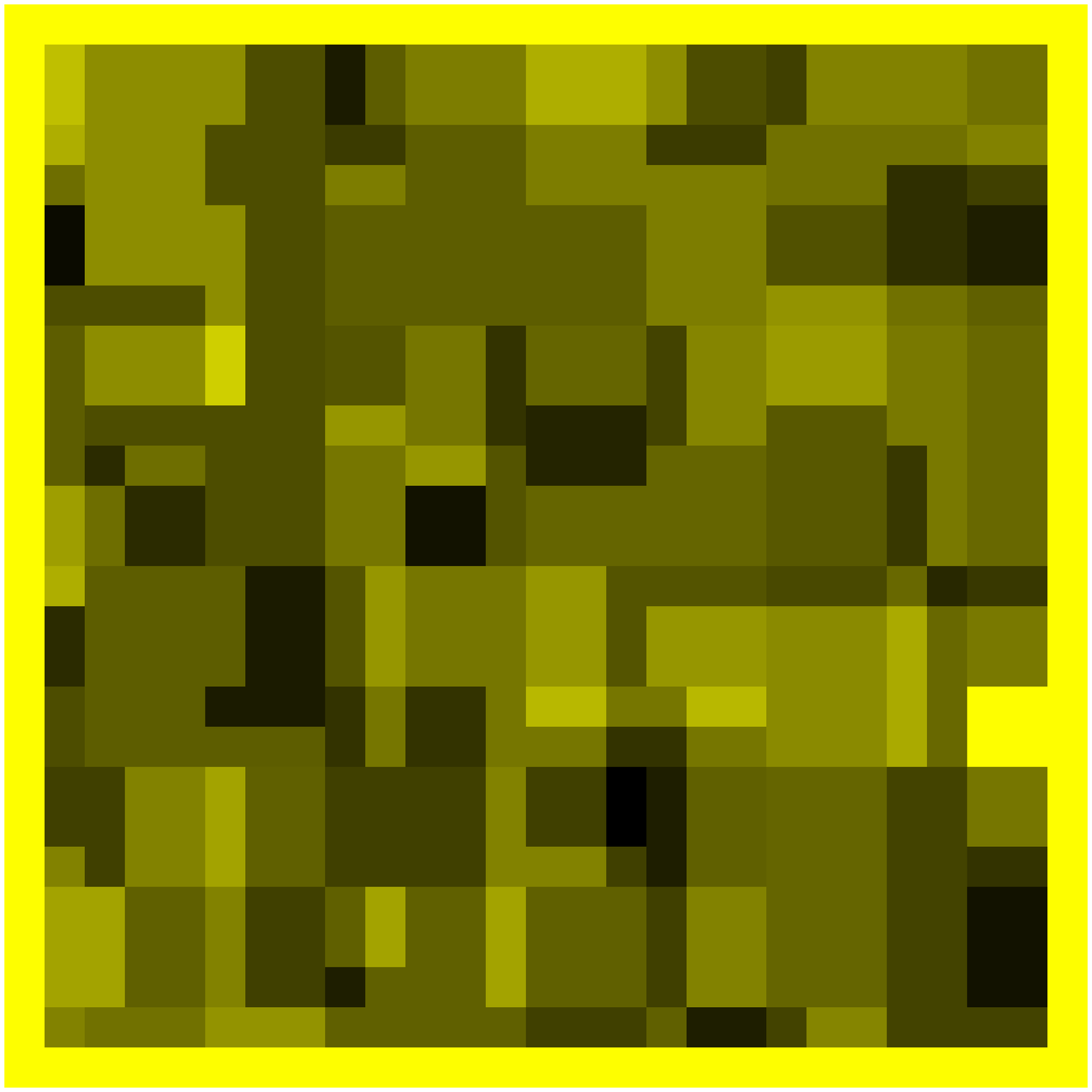}
\end{minipage}
\begin{minipage}{.1\textwidth}\centering
\includegraphics[width=1.0\textwidth]{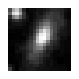}
\end{minipage}
\begin{minipage}{.1\textwidth}\centering
\includegraphics[width=1.0\textwidth]{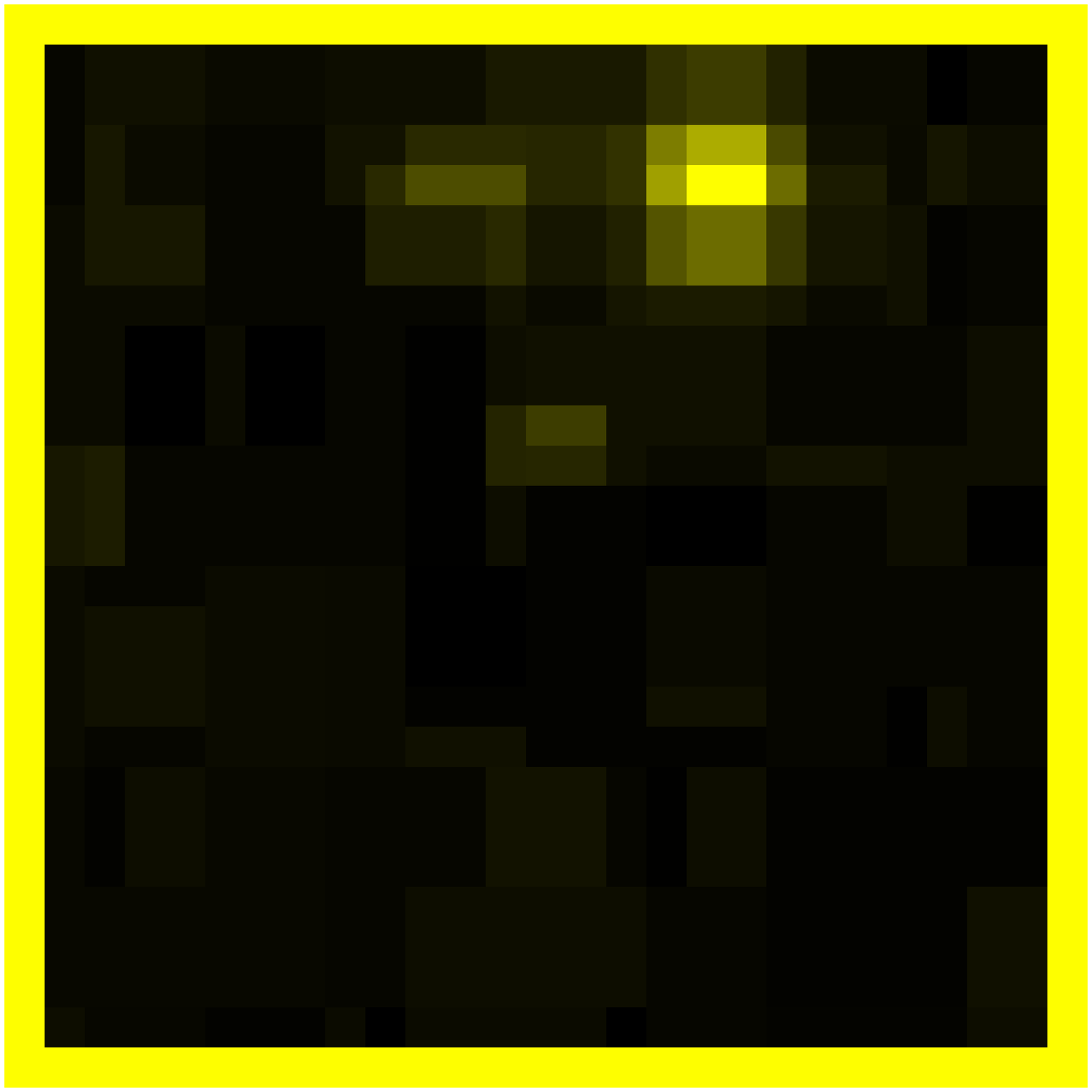}
\end{minipage}
\begin{minipage}{.1\textwidth}\centering
\includegraphics[width=1.0\textwidth]{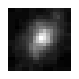}
\end{minipage}
\begin{minipage}{.1\textwidth}\centering
\includegraphics[width=1.0\textwidth]{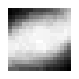}
\end{minipage}
\begin{minipage}{.1\textwidth}\centering
\includegraphics[width=1.0\textwidth]{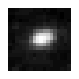}
\end{minipage}
\begin{minipage}{.1\textwidth}\centering
\includegraphics[width=1.0\textwidth]{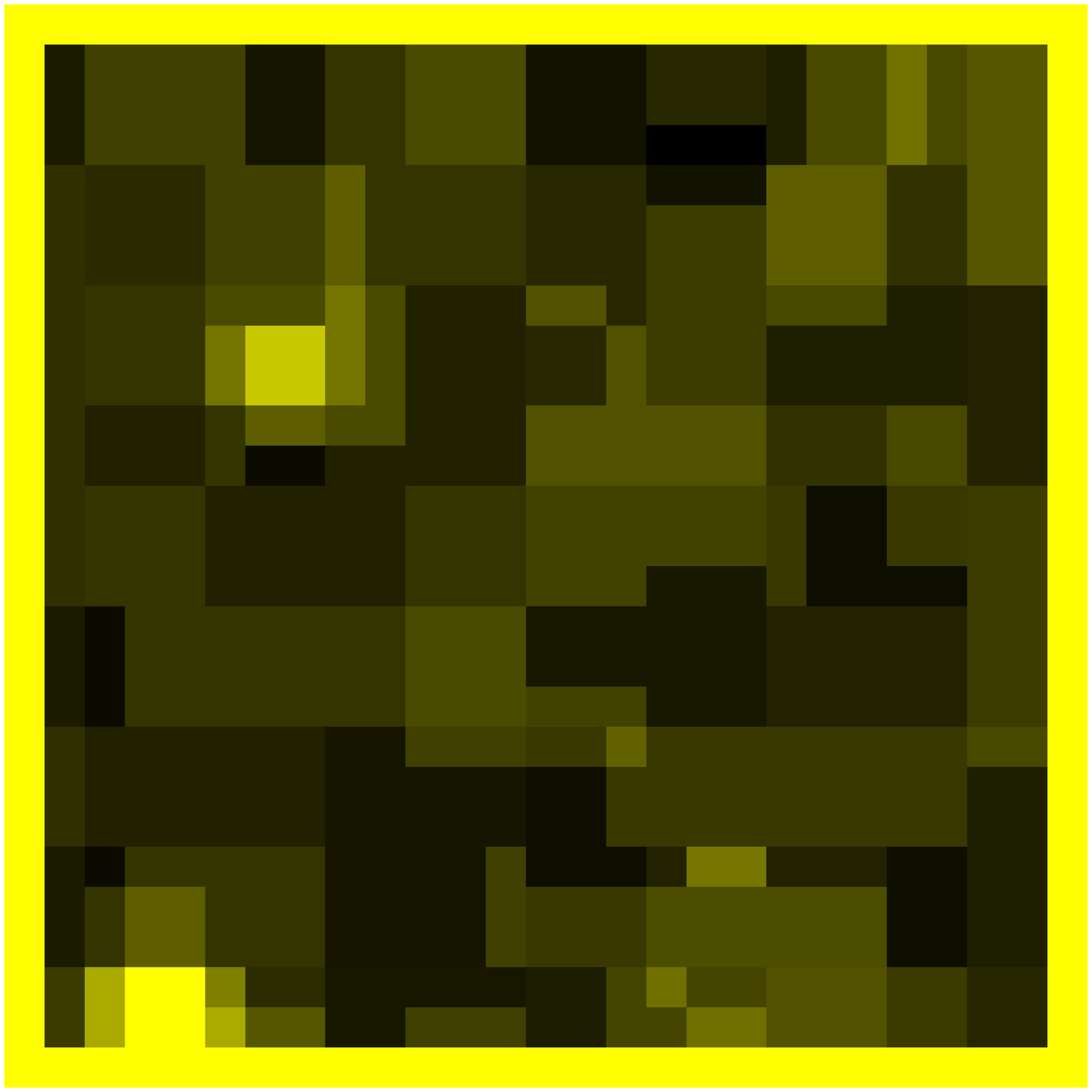}
\end{minipage}
\begin{minipage}{.1\textwidth}\centering
\includegraphics[width=1.0\textwidth]{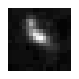}
\end{minipage}
\begin{minipage}{.1\textwidth}\centering
\includegraphics[width=1.0\textwidth]{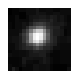}
\end{minipage}
\begin{minipage}{.1\textwidth}\centering
\includegraphics[width=1.0\textwidth]{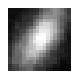}
\end{minipage}
\begin{minipage}{.1\textwidth}\centering
\includegraphics[width=1.0\textwidth]{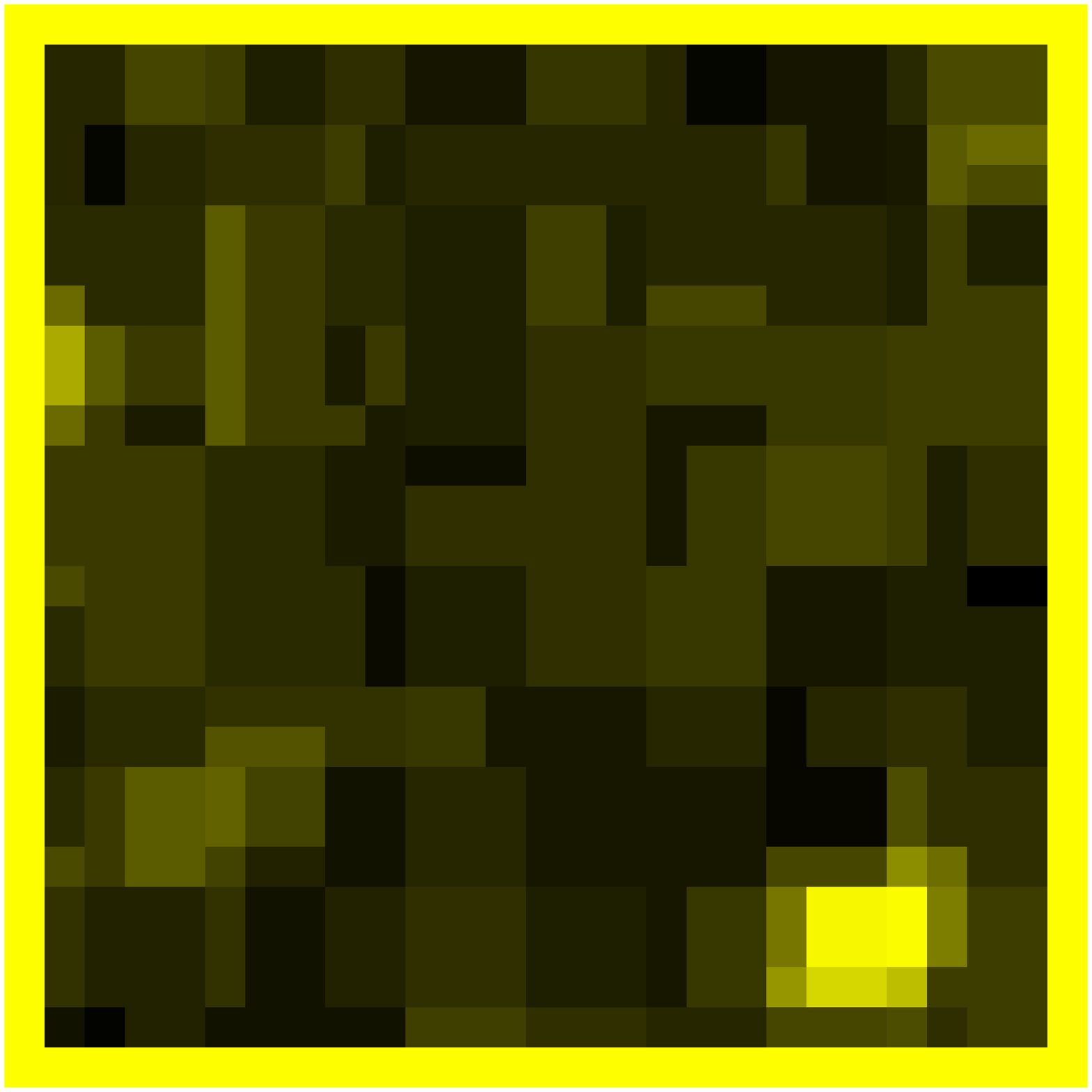}
\end{minipage}
\begin{minipage}{.1\textwidth}\centering
\includegraphics[width=1.0\textwidth]{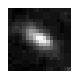}
\end{minipage}
\begin{minipage}{.1\textwidth}\centering
\includegraphics[width=1.0\textwidth]{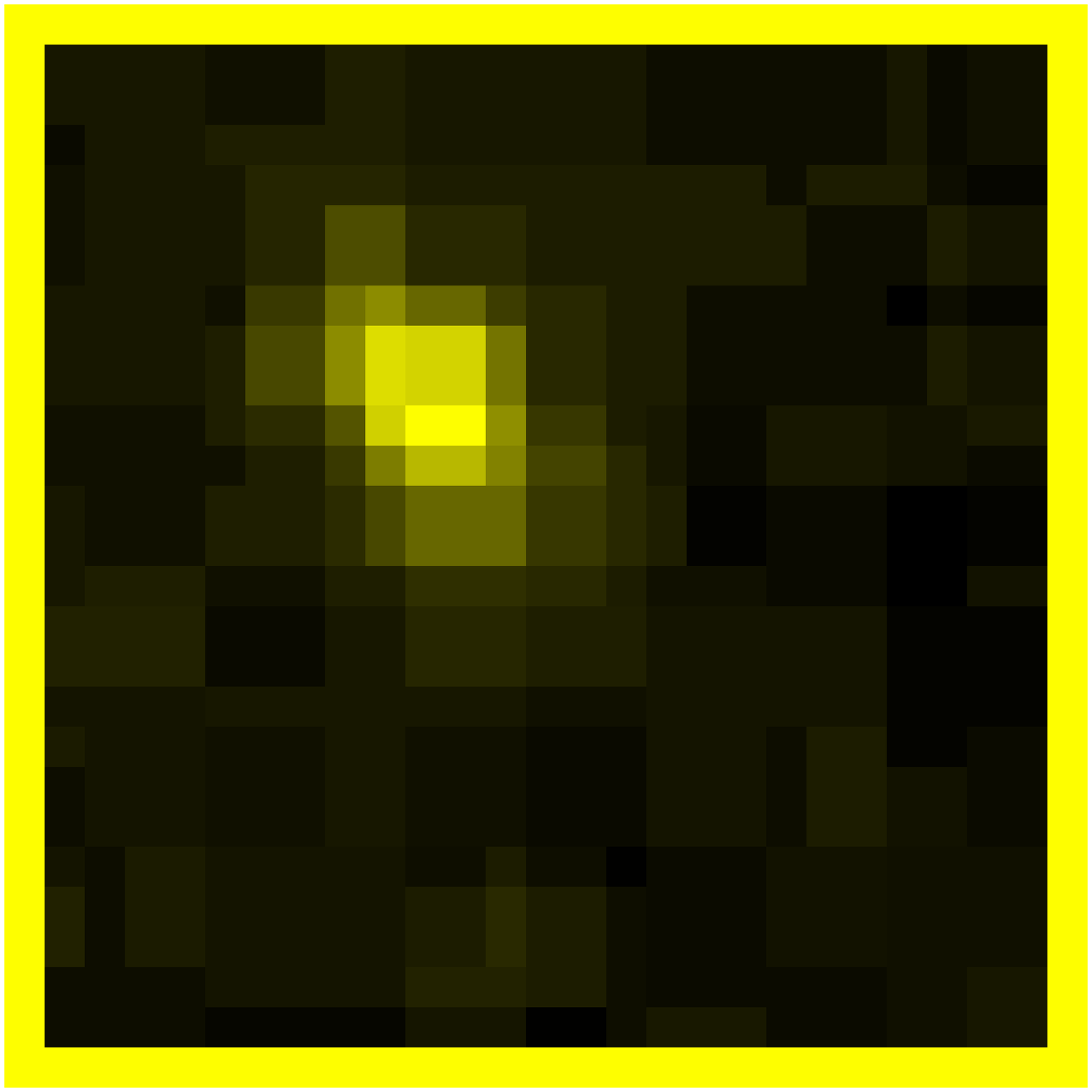}
\end{minipage}
\begin{minipage}{.1\textwidth}\centering
\includegraphics[width=1.0\textwidth]{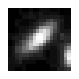}
\end{minipage}
\begin{minipage}{.1\textwidth}\centering
\includegraphics[width=1.0\textwidth]{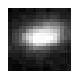}
\end{minipage}
\begin{minipage}{.1\textwidth}\centering
\includegraphics[width=1.0\textwidth]{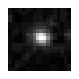}
\end{minipage}
\caption{DSS images (image size=$0.5\times0.5$ arcmin$^2$; image scale=linear) of the IRAS sources in Cat. (I), centred at the IRAS position of each source. The images are ordered in the same way as in Table~\ref{catI} from left to right, top to bottom.}\label{fig:DSSimage}
\end{figure*}

\subsection{$90$ micron flux interpolation}
The AKARI WIDE-S band channel has a 90\,$\mu$m effective wavelength. To estimate 90\,$\mu$m flux, we need to interpolate using IRAS flux measurements at 60 and 100\,$\mu$m ($F_{60}^{IRAS}$ and $F_{100}^{IRAS}$). According to the IRAS 100\,$\mu$m flux quality, we divide our objects, all of which have high quality $F_{60}^{IRAS}$, into two categories: (I) moderate quality or (II) upper limit (i.e. undetected). One caveat to bear in mind is that IRAS flux densities with moderate quality are not highly reliable as they do not satisfy the reliability requirement of the FSC which is $\geq 90\%$ at 12, 25\, $\mu$m and $\geq 80\%$ at 60\,$\mu$m. Moreover, moderate quality $F_{100}^{IRAS}$ may be caused by cirrus and not the source itself (Moshir et al. 1992). 

Consequently, we find 23 IRAS sources in Cat. (I). The interpolated 90\,$\mu$m flux $F_{90}^{int}$ is obtained by fitting $F_{60}^{IRAS}$ and $F_{100}^{IRAS}$ with a $\nu$ $\times$ $B_{\nu}(T)$ curve based on the far-infrared spectrum derived from radiative transfer models for cirrus and starburst galaxies. In Table \ref{catI}, we present source name, non-colour corrected $F_{60}^{IRAS}$, $F_{100}^{IRAS}$ and $F_{90}^{int}$, redshift (if available) and observed wavebands for each source. The average colour at 90-60 $\mu m$ $\langle F_{90}/F_{60} \rangle$ is 2.48. Colour correction is needed for SEDs other than $F_{\nu} \propto \frac{1}{\nu}$. For ordinary stars and galaxies, this correction is of order a few percent or even less. Therefore, due to the relatively large uncertainty in the flux interpolation and the peak timeline signal as a flux indicator for point sources, colour correction is neglected in this paper.
  
\begin{figure*}
\begin{minipage}{.1\textwidth}\centering
\includegraphics[width=1.0\textwidth]{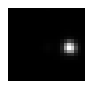}
\end{minipage}
\begin{minipage}{.1\textwidth}\centering
\includegraphics[width=1.0\textwidth]{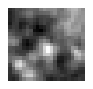}
\end{minipage}
\begin{minipage}{.1\textwidth}\centering
\includegraphics[width=1.0\textwidth]{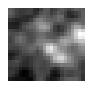}
\end{minipage}
\begin{minipage}{.1\textwidth}\centering
\includegraphics[width=1.0\textwidth]{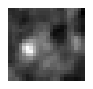}
\end{minipage}
\begin{minipage}{.1\textwidth}\centering
\includegraphics[width=1.0\textwidth]{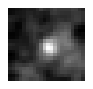}
\end{minipage}
\begin{minipage}{.1\textwidth}\centering
\includegraphics[width=1.0\textwidth]{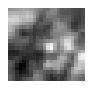}
\end{minipage}
\begin{minipage}{.1\textwidth}\centering
\includegraphics[width=1.0\textwidth]{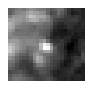}
\end{minipage}
\begin{minipage}{.1\textwidth}\centering
\includegraphics[width=1.0\textwidth]{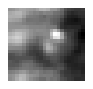}
\end{minipage}
\begin{minipage}{.1\textwidth}\centering
\includegraphics[width=1.0\textwidth]{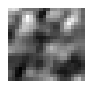}
\end{minipage}
\begin{minipage}{.1\textwidth}\centering
\includegraphics[width=1.0\textwidth]{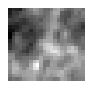}
\end{minipage}
\begin{minipage}{.1\textwidth}\centering
\includegraphics[width=1.0\textwidth]{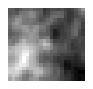}
\end{minipage}
\begin{minipage}{.1\textwidth}\centering
\includegraphics[width=1.0\textwidth]{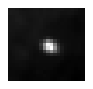}
\end{minipage}
\begin{minipage}{.1\textwidth}\centering
\includegraphics[width=1.0\textwidth]{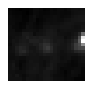}
\end{minipage}
\begin{minipage}{.1\textwidth}\centering
\includegraphics[width=1.0\textwidth]{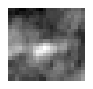}
\end{minipage}
\begin{minipage}{.1\textwidth}\centering
\includegraphics[width=1.0\textwidth]{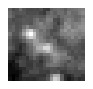}
\end{minipage}
\begin{minipage}{.1\textwidth}\centering
\includegraphics[width=1.0\textwidth]{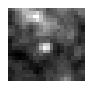}
\end{minipage}
\begin{minipage}{.1\textwidth}\centering
\includegraphics[width=1.0\textwidth]{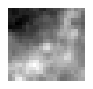}
\end{minipage}
\begin{minipage}{.1\textwidth}\centering
\includegraphics[width=1.0\textwidth]{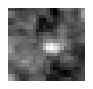}
\end{minipage}
\begin{minipage}{.1\textwidth}\centering
\includegraphics[width=1.0\textwidth]{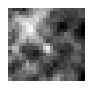}
\end{minipage}
\begin{minipage}{.1\textwidth}\centering
\includegraphics[width=1.0\textwidth]{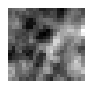}
\end{minipage}
\begin{minipage}{.1\textwidth}\centering
\includegraphics[width=1.0\textwidth]{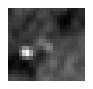}
\end{minipage}
\begin{minipage}{.1\textwidth}\centering
\includegraphics[width=1.0\textwidth]{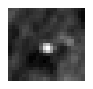}
\end{minipage}
\begin{minipage}{.1\textwidth}\centering
\includegraphics[width=1.0\textwidth]{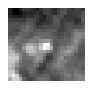}
\end{minipage}
\caption{IRAS 60\,$\mu$m images (image size=$0.5\times0.5$ deg.$^2$; image scale=linear) of the IRAS sources in Cat. (I), centred at the IRAS position of each source and ordered in the same way as in Fig.~\ref{fig:DSSimage}.}\label{fig:IRAS60image}
\end{figure*}

\begin{figure*}
\begin{minipage}{.1\textwidth}\centering
\includegraphics[width=1.0\textwidth]{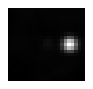}
\end{minipage}
\begin{minipage}{.1\textwidth}\centering
\includegraphics[width=1.0\textwidth]{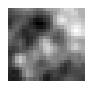}
\end{minipage}
\begin{minipage}{.1\textwidth}\centering
\includegraphics[width=1.0\textwidth]{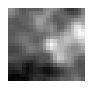}
\end{minipage}
\begin{minipage}{.1\textwidth}\centering
\includegraphics[width=1.0\textwidth]{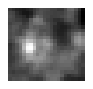}
\end{minipage}
\begin{minipage}{.1\textwidth}\centering
\includegraphics[width=1.0\textwidth]{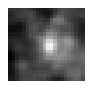}
\end{minipage}
\begin{minipage}{.1\textwidth}\centering
\includegraphics[width=1.0\textwidth]{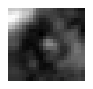}
\end{minipage}
\begin{minipage}{.1\textwidth}\centering
\includegraphics[width=1.0\textwidth]{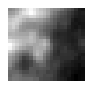}
\end{minipage}
\begin{minipage}{.1\textwidth}\centering
\includegraphics[width=1.0\textwidth]{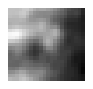}
\end{minipage}
\begin{minipage}{.1\textwidth}\centering
\includegraphics[width=1.0\textwidth]{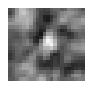}
\end{minipage}
\begin{minipage}{.1\textwidth}\centering
\includegraphics[width=1.0\textwidth]{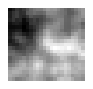}
\end{minipage}
\begin{minipage}{.1\textwidth}\centering
\includegraphics[width=1.0\textwidth]{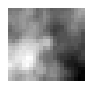}
\end{minipage}
\begin{minipage}{.1\textwidth}\centering
\includegraphics[width=1.0\textwidth]{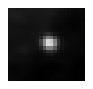}
\end{minipage}
\begin{minipage}{.1\textwidth}\centering
\includegraphics[width=1.0\textwidth]{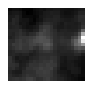}
\end{minipage}
\begin{minipage}{.1\textwidth}\centering
\includegraphics[width=1.0\textwidth]{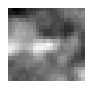}
\end{minipage}
\begin{minipage}{.1\textwidth}\centering
\includegraphics[width=1.0\textwidth]{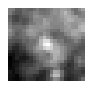}
\end{minipage}
\begin{minipage}{.1\textwidth}\centering
\includegraphics[width=1.0\textwidth]{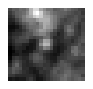}
\end{minipage}
\begin{minipage}{.1\textwidth}\centering
\includegraphics[width=1.0\textwidth]{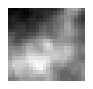}
\end{minipage}
\begin{minipage}{.1\textwidth}\centering
\includegraphics[width=1.0\textwidth]{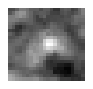}
\end{minipage}
\begin{minipage}{.1\textwidth}\centering
\includegraphics[width=1.0\textwidth]{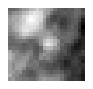}
\end{minipage}
\begin{minipage}{.1\textwidth}\centering
\includegraphics[width=1.0\textwidth]{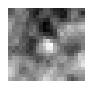}
\end{minipage}
\begin{minipage}{.1\textwidth}\centering
\includegraphics[width=1.0\textwidth]{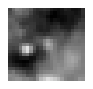}
\end{minipage}
\begin{minipage}{.1\textwidth}\centering
\includegraphics[width=1.0\textwidth]{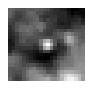}
\end{minipage}
\begin{minipage}{.1\textwidth}\centering
\includegraphics[width=1.0\textwidth]{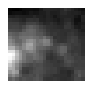}
\end{minipage}
\caption{IRAS 100\,$\mu$m images (image size=$0.5\times0.5$ deg.$^2$; image scale=linear) of the IRAS sources in Cat. (I), centred at the IRAS position of each source and ordered in the same way as in Fig.~\ref{fig:DSSimage}.}\label{fig:IRAS100image}
\end{figure*}

\begin{figure*}
\begin{minipage}{.1\textwidth}\centering
\includegraphics[width=1.0\textwidth]{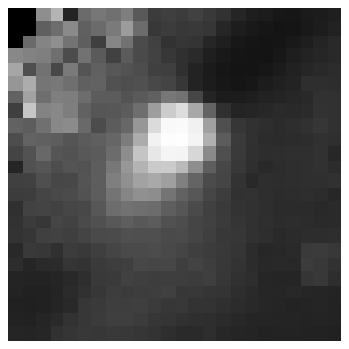}
\end{minipage}
\begin{minipage}{.1\textwidth}\centering
\includegraphics[width=1.0\textwidth]{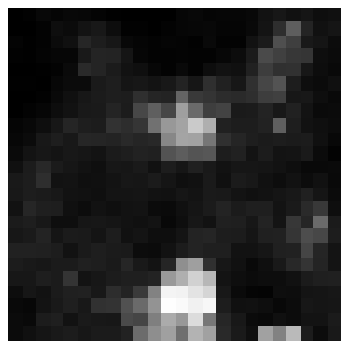}
\end{minipage}
\begin{minipage}{.1\textwidth}\centering
\includegraphics[width=1.0\textwidth]{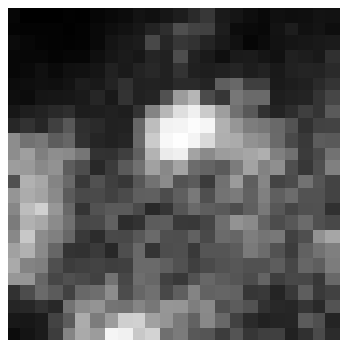}
\end{minipage}
\begin{minipage}{.1\textwidth}\centering
\includegraphics[width=1.0\textwidth]{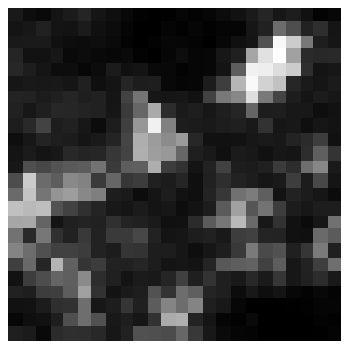}
\end{minipage}
\begin{minipage}{.1\textwidth}\centering
\includegraphics[width=1.0\textwidth]{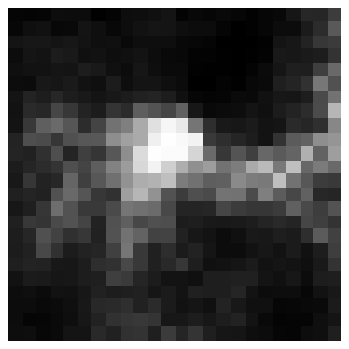}
\end{minipage}
\begin{minipage}{.1\textwidth}\centering
\includegraphics[width=1.0\textwidth]{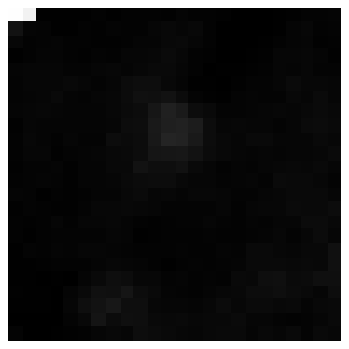}
\end{minipage}
\begin{minipage}{.1\textwidth}\centering
\includegraphics[width=1.0\textwidth]{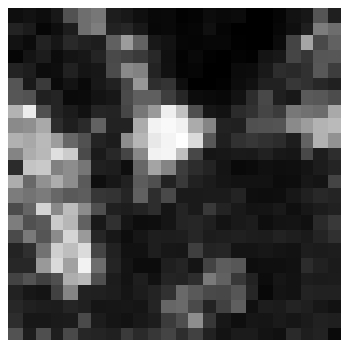}
\end{minipage}
\begin{minipage}{.1\textwidth}\centering
\includegraphics[width=1.0\textwidth]{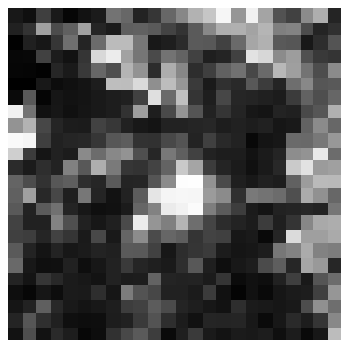}
\end{minipage}
\begin{minipage}{.1\textwidth}\centering
\includegraphics[width=1.0\textwidth]{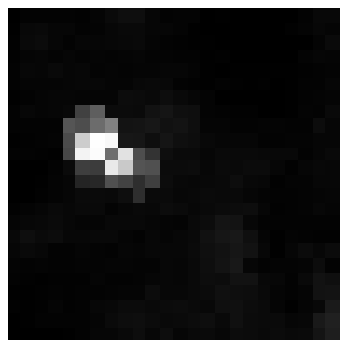}
\end{minipage}
\begin{minipage}{.1\textwidth}\centering
\includegraphics[width=1.0\textwidth]{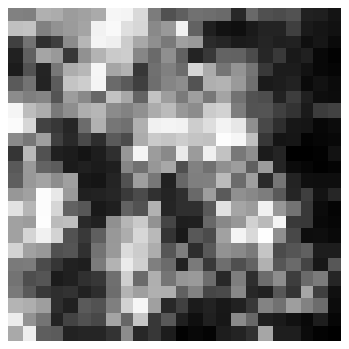}
\end{minipage}
\begin{minipage}{.1\textwidth}\centering
\includegraphics[width=1.0\textwidth]{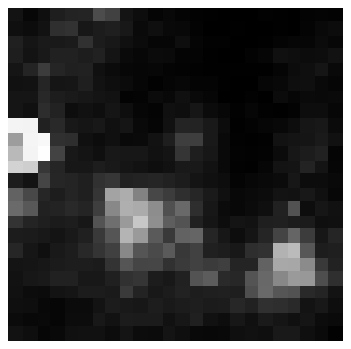}
\end{minipage}
\begin{minipage}{.1\textwidth}\centering
\includegraphics[width=1.0\textwidth]{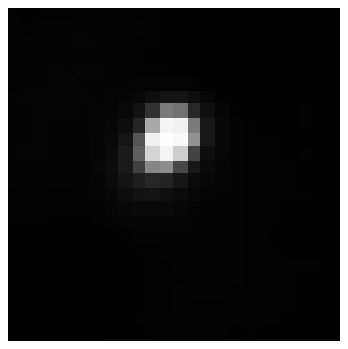}
\end{minipage}
\begin{minipage}{.1\textwidth}\centering
\includegraphics[width=1.0\textwidth]{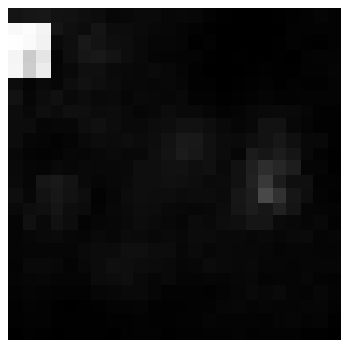}
\end{minipage}
\begin{minipage}{.1\textwidth}\centering
\includegraphics[width=1.0\textwidth]{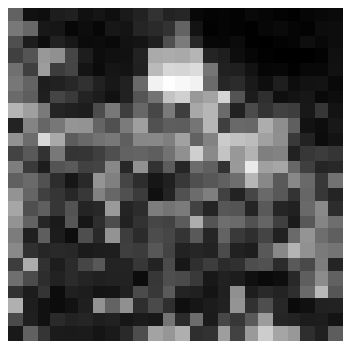}
\end{minipage}
\begin{minipage}{.1\textwidth}\centering
\includegraphics[width=1.0\textwidth]{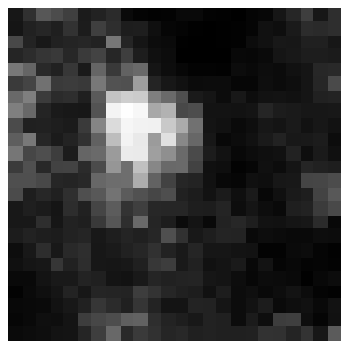}
\end{minipage}
\begin{minipage}{.1\textwidth}\centering
\includegraphics[width=1.0\textwidth]{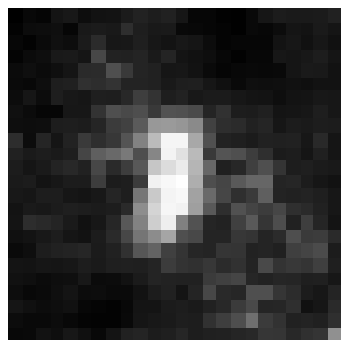}
\end{minipage}
\begin{minipage}{.1\textwidth}\centering
\includegraphics[width=1.0\textwidth]{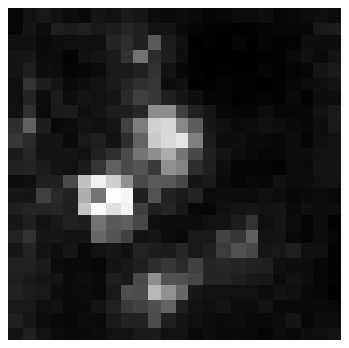}
\end{minipage}
\begin{minipage}{.1\textwidth}\centering
\includegraphics[width=1.0\textwidth]{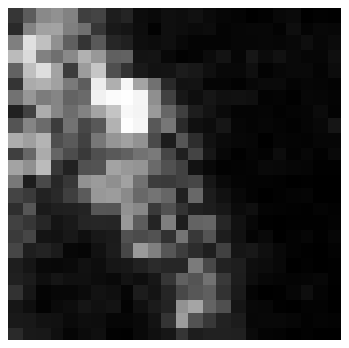}
\end{minipage}
\begin{minipage}{.1\textwidth}\centering
\includegraphics[width=1.0\textwidth]{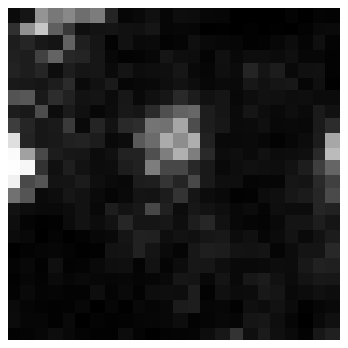}
\end{minipage}
\begin{minipage}{.1\textwidth}\centering
\includegraphics[width=1.0\textwidth]{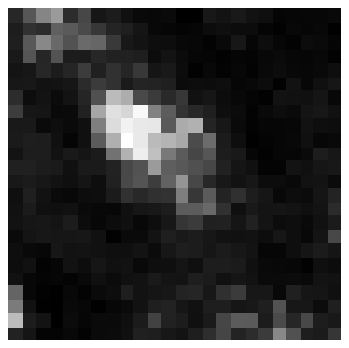}
\end{minipage}
\begin{minipage}{.1\textwidth}\centering
\includegraphics[width=1.0\textwidth]{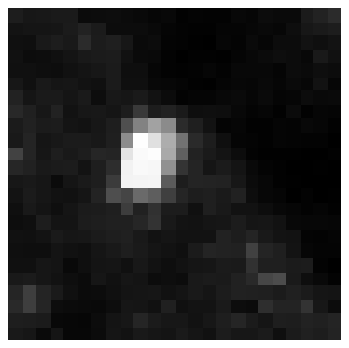}
\end{minipage}
\begin{minipage}{.1\textwidth}\centering
\includegraphics[width=1.0\textwidth]{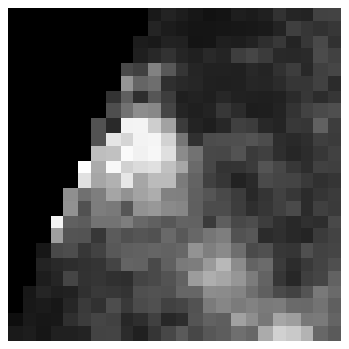}
\end{minipage}
\begin{minipage}{.1\textwidth}\centering
\includegraphics[width=1.0\textwidth]{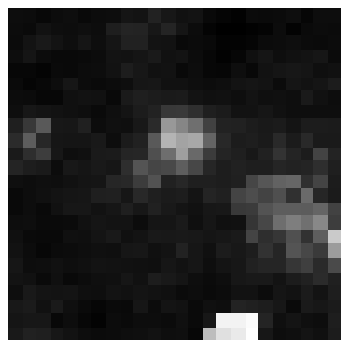}
\end{minipage}
\caption{AKARI 90\,$\mu$m images (image size=$0.1\times0.1$ deg.$^2$; image scale=squared) of the IRAS sources in Cat. (I), centred at the IRAS position of each source and ordered in the same way as in Fig.~\ref{fig:DSSimage}.}\label{fig:SXTimage}
\end{figure*}

To further check the reliability of the IRAS sources in Cat. (I), we have used the optical counterpart identification method. Fig.~\ref{fig:DSSimage} shows postage stamps from the Digitized Sky Survey (DSS) centred at the IRAS position of each source. About 6 IRAS sources (shown in yellow colour) do not have obvious counterparts present on the optical photographs. From Table~\ref{catI}, we find that these sources (with * in front of the source name) have only been observed at 60 and 100\,$\mu$m and their 60\,$\mu$m flux densities are $\sim$ 0.2 Jy, i.e. they are very faint objects. However, all IRAS sources in Cat. (I) appear to be reliable extractions judging from the IRAS $60$ and 100\,$\mu$m maps (Fig. \ref{fig:IRAS60image} and Fig. \ref{fig:IRAS100image}). The IRAS source F18001+6636 seems to be missing in these maps because of the adjacent Cat's Eye Nebula. Fig. \ref{fig:SXTimage} shows the co-added AKARI 90\,$\mu$m images centred at the IRAS positions of each source in Cat. (I). In Section \ref{detectionFSC}, we will derive the conversion factor by comparing $F_{90}^{int}$ with the AKARI peak timeline signal for each IRAS source in Cat. (I). 

For IRAS sources in Cat. (II), i.e. sources undetected at 100\,$\mu$m, we can predict 90\,$\mu$m flux density by adopting $\langle F_{100}/F_{60} \rangle =2.24$, which is derived from the IRAS Faint Source Survey (FSS) redshift survey of 700 square degrees. Although IRAS sources in Cat. (II) will not used in calibration, it is interesting to compare the predicted flux densities with the AKARI measurements.

\begin{table*}
\caption{A total of 23 IRAS sources in Cat. (I).}\label{catI}
\begin{tabular}[pos]{rllllr}
Name&$F_{60}^{IRAS} (Jy)$&$F_{90}^{int} (Jy)$&$F_{100}^{IRAS}$ (Jy)&Redshift&Observed Wavebands\\
\hline
F18001+6636 & 2.04E+00 & 2.31E+00  & 2.17E+00 & 0.026&B, 2MASS ($J,H,K_s$),IRAS(12,25,60,100 $\mu m$), ...\\
*F18112+6503 & 2.10E-01 & 5.94E-01  & 6.65E-01  & -&IRAS (60,100 $\mu m$)\\
F18130+6455 & 2.43E-01 & 6.34E-01  & 6.99E-01  & -&2MASS ($J,H,K_s$), IRAS (12,60,100 $\mu m$)\\
F18185+6341 & 2.04E-01 & 6.53E-01  & 7.49E-01  & -&2MASS ($J,H,K_s$), IRAS (60,100 $\mu m$)\\
F18197+6339 & 6.01E-01 & 1.02E+00  & 1.04E+00  & 0.027&2MASS ($J,H,K_s$),IRAS(12,25,60,100 $\mu m$), 170 $\mu m$, ...\\
F18252+6315 & 3.89E-01 & 9.14E-01  & 9.87E-01 & 0.084&2MASS ($J,H,K_s$), IRAS (25,60,100 $\mu m$)\\
F18286+6309 & 7.72E-01 & 8.33E-01  & 7.74E-01 & -&IRAS (25,60,100 $\mu m$)\\
*F18293+6304 & 2.05E-01 & 5.93E-01  & 6.67E-01 & -&IRAS (60,100 $\mu m$)\\
F18223+6255 & 1.59E-01 & 5.42E-01  & 6.30E-01  & -&2MASS ($J,H,K_s$), IRAS (25,60,100 $\mu m$)\\
*F18254+6200 & 1.72E-01 & 5.95E-01  & 6.93E-01  & -&IRAS (60,100 $\mu m$)\\
F18301+6138 & 2.40E-01 & 6.45E-01  & 7.15E-01 & -&2MASS ($J,H,K_s$), IRAS (60,100 $\mu m$)\\
F18425+6036 & 1.01E+01 & 1.93E+01  & 2.01E+01 & 0.013&2MASS ($J,H,K_s$),IRAS(12,25,60,100 $\mu m$), ...\\
F18353+5950 & 5.74E-01 & 1.17E+00  & 1.24E+00 & 0.028&2MASS ($J,H,K_s$),IRAS(25,60,100 $\mu m$), ...\\
*F18436+5931 & 2.11E-01 & 7.80E-01  & 9.21E-01  & -&IRAS (60,100 $\mu m$)\\
F18440+5900 & 3.37E-01 & 8.79E-01  & 9.69E-01  & -&2MASS ($J,H,K_s$), IRAS (25,60,100 $\mu m$)\\
F18436+5847 & 5.69E-01 & 1.04E+00  & 1.08E+00 & -&2MASS ($J,H,K_s$), IRAS (25,60,100 $\mu m$)\\
F18506+5801 & 3.79E-01 & 9.12E-01  & 9.89E-01 & 0.029&B, 2MASS ($J,H,K_s$),IRAS(25,60,100 $\mu m$), 170 $\mu m$, ...\\
*F18520+5715 & 2.11E-01 & 7.27E-01  & 8.46E-01  & -&IRAS (60,100 $\mu m$)\\
F19026+5654 & 2.99E-01 & 7.23E-01  & 7.85E-01 & -&2MASS ($J,H,K_s$), IRAS (25,60,100 $\mu m$)\\
*F18597+5631 & 1.64E-01 & 5.43E-01  & 6.27E-01  & -&IRAS (60,100 $\mu m$)\\
F18499+5542 & 4.28E-01 & 1.00E+00  & 1.08E+00  & -&IRAS (60,100 $\mu m$), ...\\
F18510+5539 & 1.25E+00 & 2.41E+00  & 2.51E+00  & 0.0242&MASS ($J,H,K_s$), IRAS (25,60,100 $\mu m$), 170 $\mu m$, ...\\
F19009+5507 & 5.02E-01 & 8.28E-01  & 8.35E-01  & -&2MASS ($J,H,K_s$), IRAS (25,60,100 $\mu m$)\\
\end{tabular}
\end{table*}

\section{TIMELINE ANALYSIS}
\label{timeline}
\subsection{Procedure outline}
Timeline Analysis, developed based on the IRAS PSC model, is a one-dimensional source detection method. It detects and outputs SCONs in each scan after deletion of signals below a certain threshold. Although the AKARI All-Sky Survey has a dedicated point source extraction and photometry software package, SUSSEXtractor (Savage \& Oliver 2007), Timeline Analysis is a valuable tool at the initial performance evaluation stage because its transparency provides insight on the timelines and guidelines on parameter tuning in the dedicated source extraction pipeline. 

Our goal is to derive the conversion factor which transforms the AKARI peak timeline signal to the interpolated monochromatic 90\,$\mu$m flux density of a source. Then we can estimate the point source detection limit by converting the standard deviation in ADU to noise level in Jansky. The following is an outline of the Timeline Analysis method.

Step 1: We prepare a source-list with positions and interpolated 90\,$\mu$m flux densities in the surveyed region, which will be used to cross-match with timeline detections later.

Step 2: We read in all TSD files in the data set and outputs SCONs above a certain threshold, e.g. S/N $> 3$ or $5$.

Step 3: SCONs detected in each TSD file is cross-matched with the source-list within a searching radius of $30$ arcseconds, in accordance with the positional accuracy of the GADS. The mean RA and DEC value of the matched SCONs for a single target source as the AKARI position and the peak timeline signal is an indicator of the AKARI 90\,$\mu$m flux density. 

Step 4: Finally, the conversion factor is computed as the ratio of the interpolated 90\,$\mu$m flux $F_{90}^{int}$ to the peak timeline signal for every seconds-confirmed detection of a particular source. Thus, IRAS sources covered by more TSD files will have more derivations of the conversion factor. The average conversion factor will be used to determine the noise level of the AKARI All-Sky Survey. 

\begin{figure*}
\includegraphics[height=3.0in,width=6.6in]{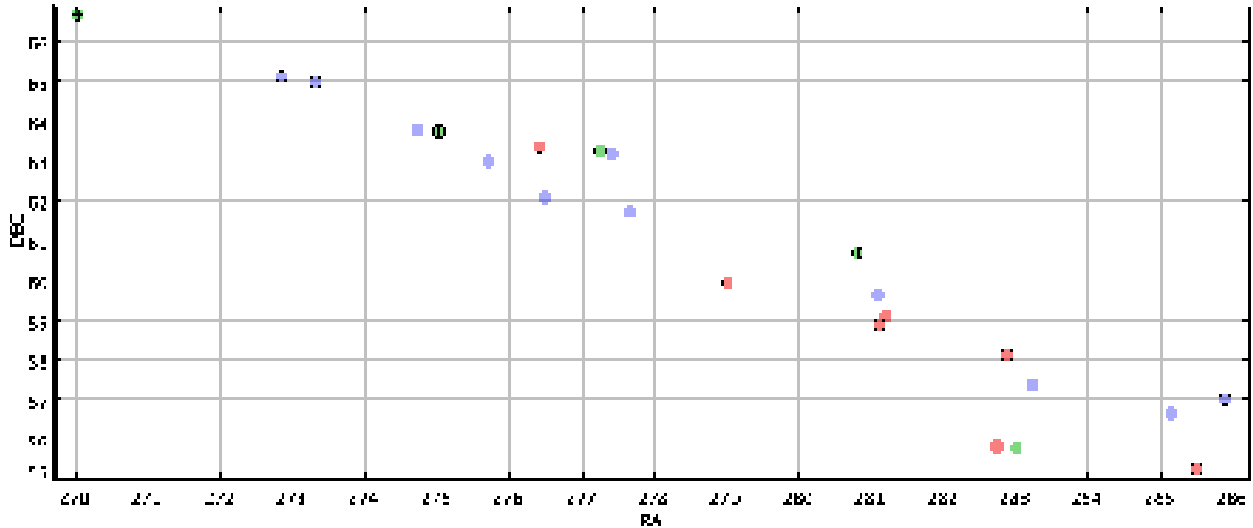}
\caption{12 seconds- and hours-confirmed timeline detections of the IRAS FSC sources in Cat. (I), thresholding at $3\sigma$ (open circles) or $5\sigma$ (plus signs). The 23 Cat. (I) sources are colour-coded according to their 60\,$\mu$m flux densities (green: $F_{60}>0.6$ Jy; red: $0.3<F_{60}<0.6$ Jy; blue: $F_{60}<0.3$ Jy).}
\label{fig:catI_5}
\end{figure*}

\subsection{Timeline source detection}
\label{detectionFSC}
Thresholding at $3\sigma$, we have detected 21 IRAS sources in Cat. (I) using the Timeline Analysis method and the two missing objects are F18293+6304 and F18254+6200. To increase the reliability of the extracted sources, we have discarded 8 sources which are only present in one TSD file\footnote{An investigation of the timelines shows that these sources are usually covered by only one TSD or at the end points of the TSD.} and 1 source whose peak timeline signals in 2 TSDs vary by more than a factor of 2. Therefore, the remaining 12 sources are F18001+6636, F18112+6503, F18130+6455, F18197+6339, F18252+6315, F18286+6309, F18425+6036, F18353+5950, F18436+5847, F18506+5801, F19026+5654 and F19009+5507. In Fig.~\ref{fig:catI_5}, we have plotted the 12 seconds- and hours-confirmed timeline detections. The IRAS FSC sources in Cat. (I) are colour-coded by the 60 micron flux measured by IRAS, $F_{60}^{IRAS}$. At S/N $>3$, nearly all sources with $F_{60}^{IRAS}>0.3$ Jy (green and red dots with open circles) as well as 2 of the faint ones (blue dots with open circles)are detected with seconds- and hours-confirmation. At S/N $>5$, 3 out of 5 sources with $F_{60}^{IRAS}>0.6$ (green dots with plus signs in the middle) are seen. Therefore, one can guess the equivalent $5\sigma$ point source detection limit at 60\,$\mu$m for a single detector row is $\sim$0.6 Jy. As the AKARI has covered almost the entire sky at least twice and each scan has $3$ detector rows, ideally the $5\sigma$ point source flux detection limit for a single scan will reach $\sim$0.2 Jy which is deeper than the detection limit of the IRAS PSC and comparable to the FSC. 

\begin{figure}
\includegraphics[height=2.5in,width=3.5in]{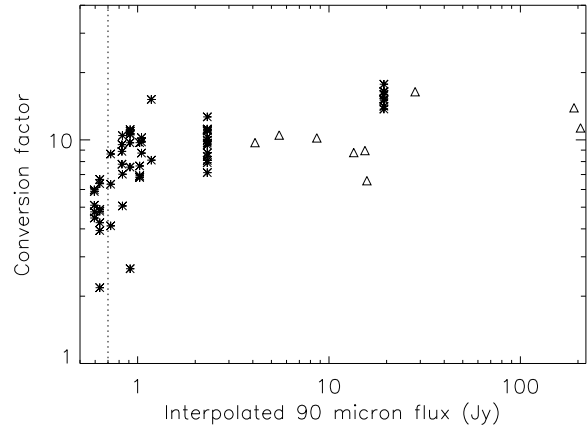}
\caption{Conversion factor ($F_{90}^{int}~/$ peak timeline signal) vs. the interpolated 90\,$\mu$m flux $F_{90}^{int}$ (asterisks: IRAS FSC sources in Cat. (I) detected at S/N $>3$; triangles: asteroids). The dotted line is where the interpolated 90\,$\mu$m flux is equal to 0.7 Jy. For clarity, error bars are not plotted.}
\label{fig:CF_IRASandASTER}
\end{figure}

Conversion factors derived from the 12 confirmed sources above $3\sigma$ and asteroids are shown in Fig.~\ref{fig:CF_IRASandASTER}. For sources with $F_{90}^{int}<0.7$ Jy (left of the dotted line in Fig.~\ref{fig:CF_IRASandASTER}), the conversion factor appears to be systematically lower than that from brighter sources. Since IRAS does not provide accurate flux determinations for these faint objects, we will exclude them in estimating the mean conversion factor. Source F18425+6036 is a nearby extended object and hence is excluded as well. So, we are left with a total of $9$ Cat. (I) IRAS sources used for absolute calibration. The mean conversion factor is computed as a weighted average,
\begin{equation}
\langle CF \rangle = \sum_{i=1}^9 \frac{CF_i}{\sigma_i^2} / \sum_{i=1}^9 \frac{1}{\sigma_i^2}  =8.49.
\end{equation}
$\sigma_i$ is the uncertainty in the conversion factor, taking into account the scatter in $F_{90}^{int}$ and the AKARI peak timeline signal. The uncertainty in the estimate of the conversion factor is calculated as 
\begin{equation}
\sigma_{\langle CF\rangle} = 1/\sqrt{\sum_{i=1}^9 \frac{1}{\sigma_i^2}} =0.61.
\end{equation}
Therefore, combined with the robust noise estimation in Section \ref{noiseest}, we deduce that in the AKARI WIDE-S 90\,$\mu$m band, the point source flux detection limit for a single detector row at S/N $>5$ is $1.1\pm0.1$ Jy (or $\sim$0.7 Jy at S/N $>3$). 

\begin{figure}
\includegraphics[height=2.5in,width=3.5in]{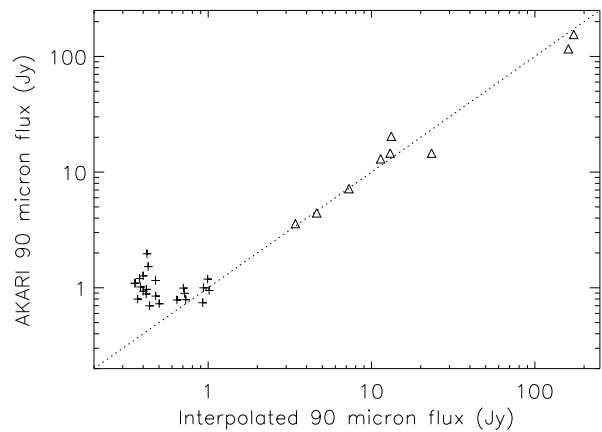}
\caption{The interpolated 90\,$\mu$m flux vs. the AKRAI 90\,$\mu$m flux for IRAS FSC sources in Cat. (II) detected at S/N $>3$ (plus signs) and asteroids (triangles).}
\label{fig:IRAS_AKARI_catII}
\end{figure}

Fig. \ref{fig:IRAS_AKARI_catII} shows a comparison between the AKARI 90\,$\mu$m flux, i.e. peak timeline signal $\times$ conversion factor, and the interpolated 90\,$\mu$m flux for Cat. (II) sources detected at a $3\sigma$ threshold. For sources with $F_{90}^{int}$ less than $\sim$0.7 Jy, the consistency between $F_{90}^{int}$ and $F_{90}^{AKARI}$ begins to break down. This trend seems to coincide with the boosting of the AKARI fluxes at the faint end present in Fig.~\ref{fig:CF_IRASandASTER}. It could be caused by the well-known flux overestimation effect from upward noise fluctuations (Moshir et al. 1992), non-linear absolute calibration, incorrect IRAS fluxes and flux interpolation (Jeong et al. 2007), or by mistaking correlated noise for faint point sources as we are approaching the detection limit of $\sim$0.7 Jy at S/N $>3$. 

\subsection{Non-IRAS sources and cirrus emission}
\label{cirrusCandi}
The expected number density of sources at $\sim$0.5 Jy is 1 source per square degree from 90\,$\mu$m integral source count models based on the European Large Area $ISO$ Survey (Efstathiou et al. 2000; Heraudeau et al. 2004). In our data set, each scan is about 3.6 degree long and 8.2 arcmin wide. In other words, each scan covers an area of $\sim$0.5 square degrees. At a $5\sigma$ threshold, two overlapping scans usually give rise to 1 - 2 seconds- and hours-confirmed sources. Therefore, with the current high source number density and sensitivity estimate, a large fraction of the detections are expected to be spurious point sources. In principle, false detections can be caused by cosmic ray glitches, transient behaviour of the detectors, correlated noise, or filamentary structures such as interstellar dust clouds or cirrus discovered by IRAS (Low et al. 1984) which dominate the background radiation in the far-infrared. Jeong et al. (2005) estimated the sky confusion noise (Helou \& Beichman 1990; Gautier et al. 1992) due to cirrus emission using simulated high-resolution dust maps for space missions such as $ISO$, Spitzer, AKARI, Herschel etc. Jeong et al. (2006) included source confusion in different source distribution models in determining the far-infrared point source detection limit. 

\begin{figure*}
\begin{minipage}{.19\textwidth}\centering
\includegraphics[width=1.0\textwidth]{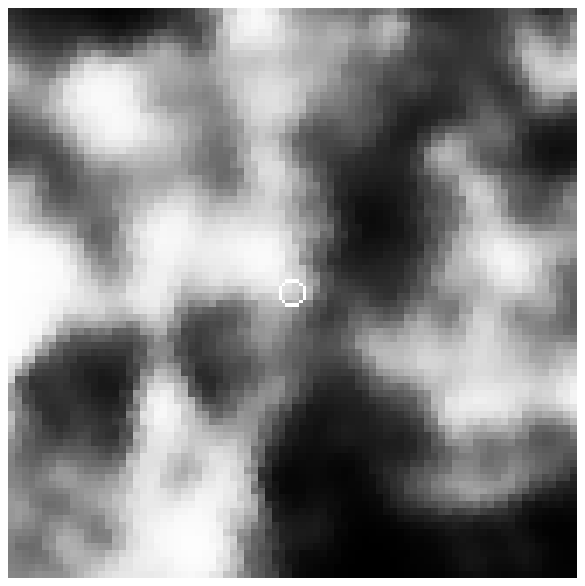}
\end{minipage}
\begin{minipage}{.19\textwidth}\centering
\includegraphics[width=1.0\textwidth]{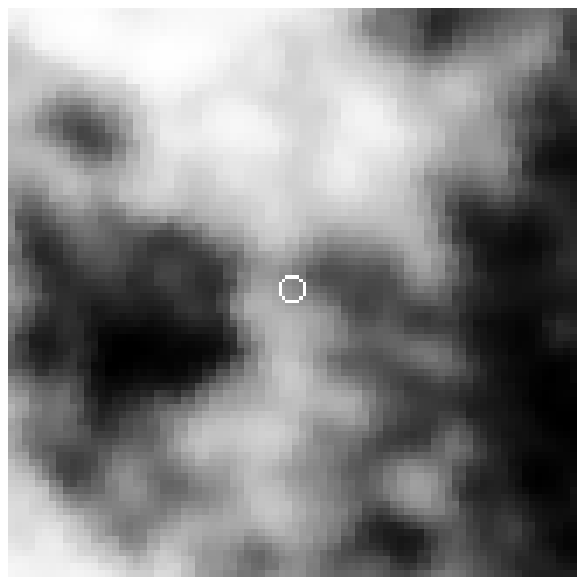}
\end{minipage}
\begin{minipage}{.19\textwidth}\centering
\includegraphics[width=1.0\textwidth]{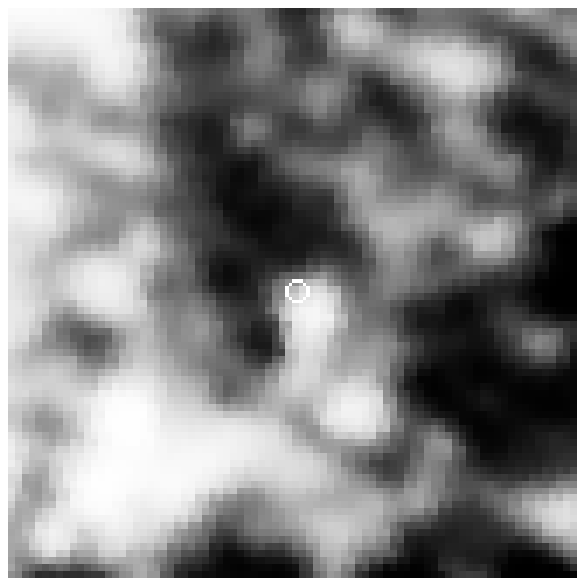}
\end{minipage}
\begin{minipage}{.19\textwidth}\centering
\includegraphics[width=1.0\textwidth]{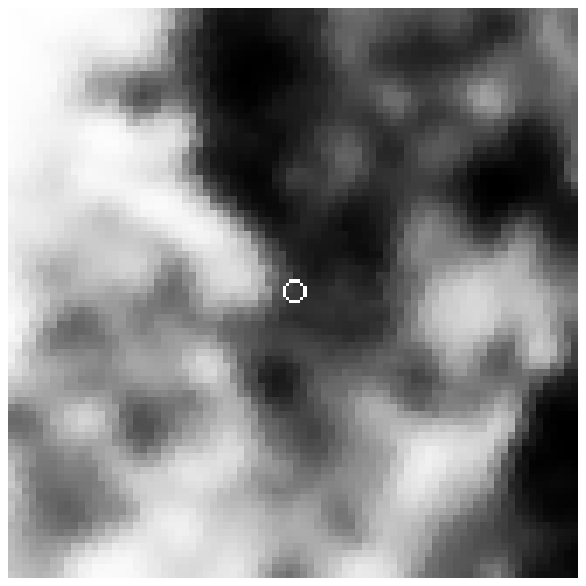}
\end{minipage}
\begin{minipage}{.19\textwidth}\centering
\includegraphics[width=1.0\textwidth]{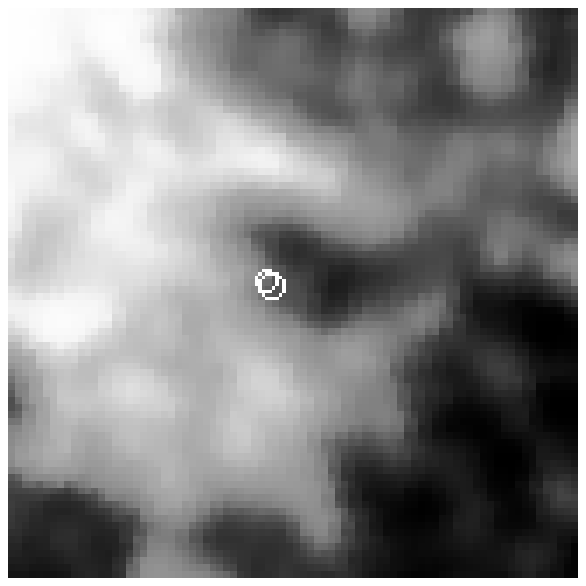}
\end{minipage}
\caption{IRAS 100\,$\mu$m maps (image size=$2\times2$ deg.$^2$; image scale=histogram). From left to right, the white open circles (diameter $=4.8$ arcminutes, close to the IRAS resolution) are the cirrus candidates C1, C2, C3, C4, C5 and C6 respectively (C5 and C6 are shown together in the rightmost map.).}\label{fig:cirrus}
\end{figure*}

\begin{figure*}
\begin{minipage}{.19\textwidth}\centering
\includegraphics[width=1.0\textwidth]{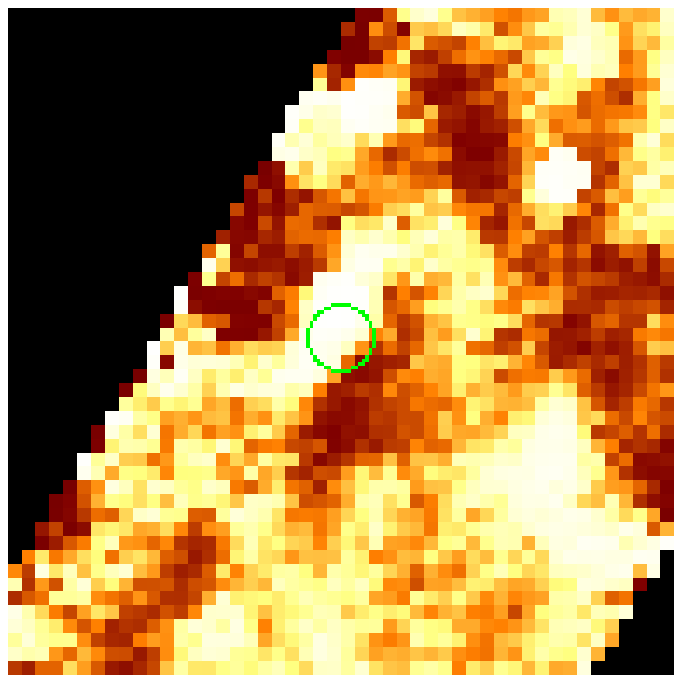}
\end{minipage}
\begin{minipage}{.19\textwidth}\centering
\includegraphics[width=1.0\textwidth]{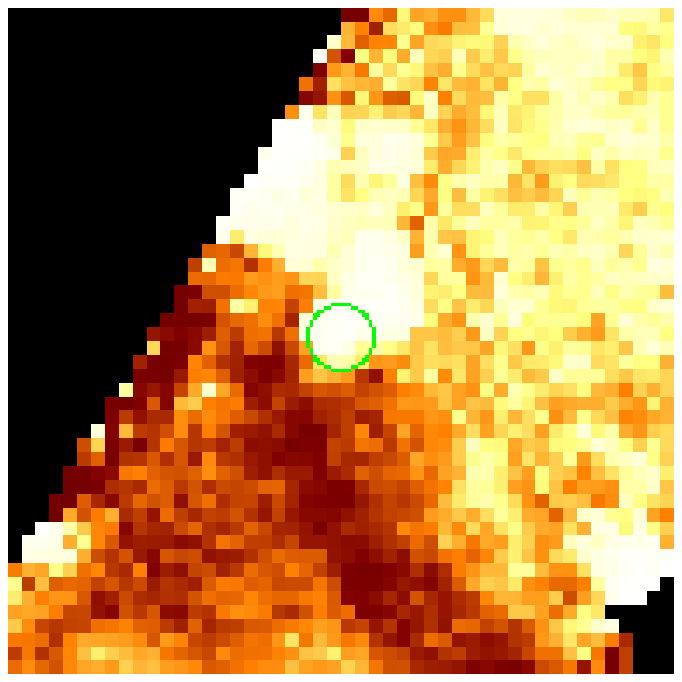}
\end{minipage}
\begin{minipage}{.19\textwidth}\centering
\includegraphics[width=1.0\textwidth]{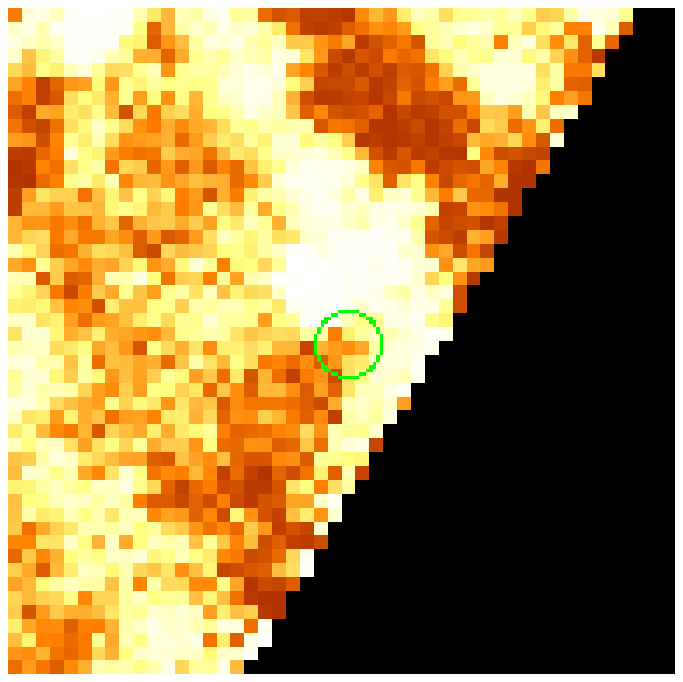}
\end{minipage}
\begin{minipage}{.19\textwidth}\centering
\includegraphics[width=1.0\textwidth]{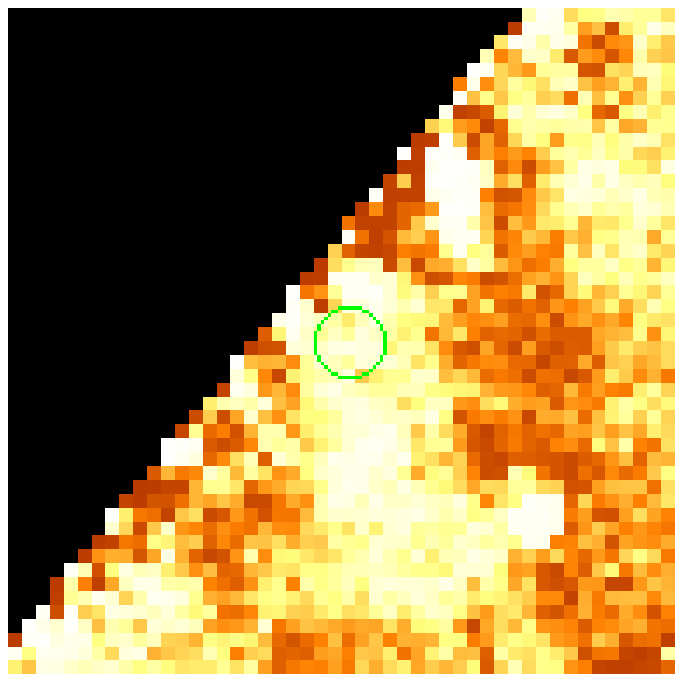}
\end{minipage}
\begin{minipage}{.19\textwidth}\centering
\includegraphics[width=1.0\textwidth]{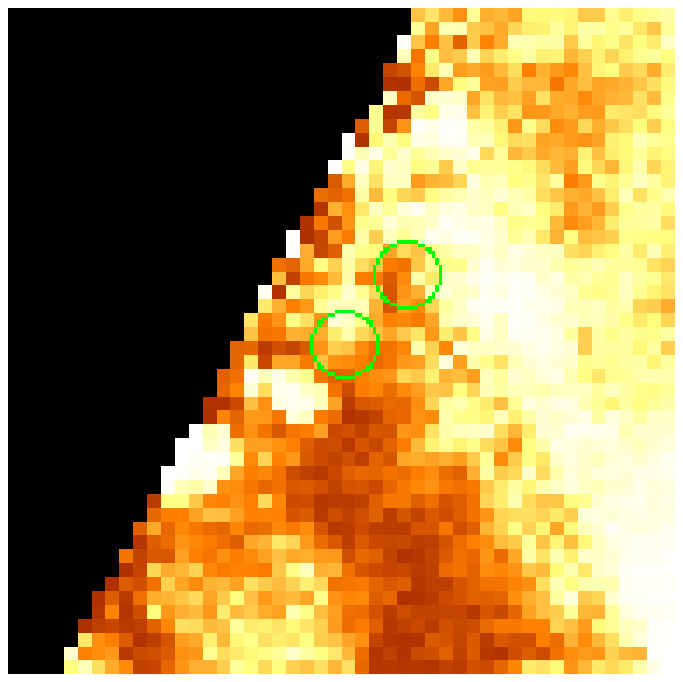}
\end{minipage}
\caption{AKARI 90\,$\mu$m maps (image size=$0.2\times0.2$ deg.$^2$; image scale=histogram). From left to right, green open circles (with diameter $=1.2$ arcminutes) are the cirrus candidates C1, C2, C3, C4, C5 and C6 respectively (C5 and C6 are shown together in the rightmost map.).}\label{fig:AKARIimage_cirrus}
\end{figure*}

To investigate the effect of infrared cirrus emission on source extraction, firstly we need to select seconds- and hours-confirmed detections with a high signal-to-noise ratio to decrease the contamination from transient signals or moving objects. Having selected $5$ pairs of overlapping scans, we output SCONs above a certain threshold (varies between $4\sigma$ and $5\sigma$) and then discard those without hours-confirmation. Thus, we have obtained a sample of 8 seconds- and hours-confirmed sources which are not IRAS sources. In Fig.~\ref{fig:cirrus}, we have overlaid $6$ of them (white open circles) on the IRAS 100\,$\mu$m emission maps (Schlegel, Finkbeiner \& Davis 1998) and found that they  generally reside in high or intermediate intensity environments. Fig.~\ref{fig:AKARIimage_cirrus} shows the co-added AKARI 90\,$\mu$m images with greater resolution and confirms the extended structures of the selected non-IRAS sources shown as green open circles (could be compared with the 90\,$\mu$m images of the point sources in Fig.~\ref{fig:SXTimage}). From now on, we refer to these objects as cirrus candidates. In Table \ref{confirmedSpurious}, we list name and position for each candidate. In Section~\ref{wavelet_cirrus}, we will use wavelet multiscale decomposition technique to verify or falsify the assumption that these seconds- and hours-confirmed non-IRAS sources are due to cirrus emission. 

For a fair comparison with the cirrus candidates selected at S/N $>4$ or $5$, we have run our timeline analysis method with a threshold of $4\sigma$. The extracted seconds- and hours-confirmed IRAS sources at S/N $>4$ are listed in Table \ref{IRAS_4sigma}.

\begin{table}
\caption{6 cirrus candidates detected at $4\sigma$ or $5\sigma$.}\label{confirmedSpurious}
\begin{tabular}[pos]{lll}
NAME   & RA (J2000)    & DEC (J2000)\\
\hline
C1     & 277.796       & 61.716\\
C2     & 286.906       & 56.427\\
C3     & 284.842       & 57.186\\
C4     & 276.261       & 63.704\\
C5     & 286.365       & 55.464\\
C6     & 286.332       & 55.483\\
\end{tabular}
\end{table}

\begin{table}
\caption{3 IRAS sources in Cat. (I) detected at $4\sigma$.}\label{IRAS_4sigma}
\begin{tabular}[pos]{lll}
NAME   & RA (J2000)    & DEC (J2000)\\
\hline
 F18001+6636 & 270.042 & 66.6118\\
 F18197+6339 & 275.029 & 63.679 \\
 F18436+5847 & 281.115 & 58.837 \\
\end{tabular}
\end{table}

\section{WAVELET TRANSFORM AND SOURCE DETECTION}
\label{wavelets}
\subsection{Key motivations}	
Infrared astronomical data are often complex combinations of random noise (photon shot noise, instrumental noise, etc.), variable sky background radiation, point and quasi-point sources (e.g. asteroids, stars, double stars, galaxies, ...), different types of glitches caused by cosmic rays, extended structures such as galaxies, clusters, cirrus clouds, etc. Different signals prefer to exhibit themselves on different scales because they originate from a hierarchy of physical structures. For instance, noise tends to dominate on small scales, infrared cirrus emission often shows up on large scales, while stars and galaxies would appear on somewhat intermediate scales. So, it is advantageous to use multiresolution techniques to detect signals from objects with a particular size. In addition, because our data always contain some discontinuities and sharp features, the traditional Fourier transform becomes an inadequate and inefficient analysis tool for modelling the real data.

For the AKARI All-Sky Survey, a multiscale analysis can effectively eliminate spurious sources if the majorities have a different spatial scale from the point source scale. Essentially, we use wavelet multiresolution decomposition as a means of understanding data characteristics and increasing the reliability and completeness of the future catalogues from the AKARI All-Sky Survey. 

\subsection{Introduction to wavelets and wavelet transforms}
\label{wavelet introduction}

Wavelet transform is in many ways similar to Fourier transform. Both are linear operators which decompose signals onto a set of basis functions. The fundamental difference is that wavelet transform adopts localised basis functions called wavelets instead of infinitely ranging sinusoidal functions which have no positional resolution at all. The mother wavelet function gives rise to a family of wavelet functions by operations of translation and scaling without changing its shape. Therefore, wavelet transform offers a time-scale two-dimensional representation of the data, i.e. an accurate local description of the frequency components present in each signal segment. It is particularly useful for analysing non-stationary signals whose characteristics change over time. Alternatively, it can be viewed as a multiresolution decomposition which is perhaps more widely used (refer to Burrus, Gopinath \& Guo 1998 for a mathematical background of wavelet analysis and Starck \& Murtagh 2002, van der Berg 2004 and references therein for applications of wavelets in astronomy and physics). 

There are two types of wavelet transform, the continuous wavelet transform (CWT) and the discrete wavelet transform (DWT). Theoretically the CWT is infinitely redundant and inefficient, however it provides a smooth wavelet transform of the signal and is suited for feature recognition. The DWT is fast to compute and the signal reconstruction is straightforward. In this paper, we use the CWT to analyse the TSDs.
 
The CWT of a square-integrable function $f(x)$ is defined as
\begin{equation}
W(a, b)=\frac{1}{\sqrt{a}} \int_{-\infty}^{+\infty} f(x) \psi^* (\frac{x-b}{a}) dx, 
\end{equation}
where $\psi^*(x)$ denotes the complex conjugate of $\psi(x)$. By the convolution theorem, the wavelet transform is the inverse Fourier transform of the product of 
$\widehat{f}(w)$ and $\widehat{\psi^*}(aw)$, 
\begin{equation}
W(a, b)=\sqrt{a} \int_{-\infty}^{+\infty} \widehat{\psi^*}(aw) \widehat{f}(w) e^{ibw} dw.
\end{equation}
The wavelet power spectrum (WPS) is defined as $|W(a,b)|^2$. Regions of large or small power in the WPS plot indicate important or negligible features within the signal.

There are several factors in determining which wavelet function to use (Torrence \& Compo 1998). Essentially, wavelet transform calculates the correlation of the wavelet function and the local signal. Therefore, to detect point sources in the timelines, we need wavelet functions which can produce large wavelet coefficients at the position of the source and at the same time reduce the effect of noise. We use the Gaussian second derivative (also known as the Marr wavelet or Mexican Hat wavelet) as our analysing wavelet function. It has been suggested by many that the optimal wavelet to detect point sources is perhaps the isotropic Mexican Hat wavelet (Cay\'{o}n et al. 2000; Gonz\'{a}lez-Nuevo et al. 2006). 

\begin{figure*}
\includegraphics[height=2.0in,width=3.2in]{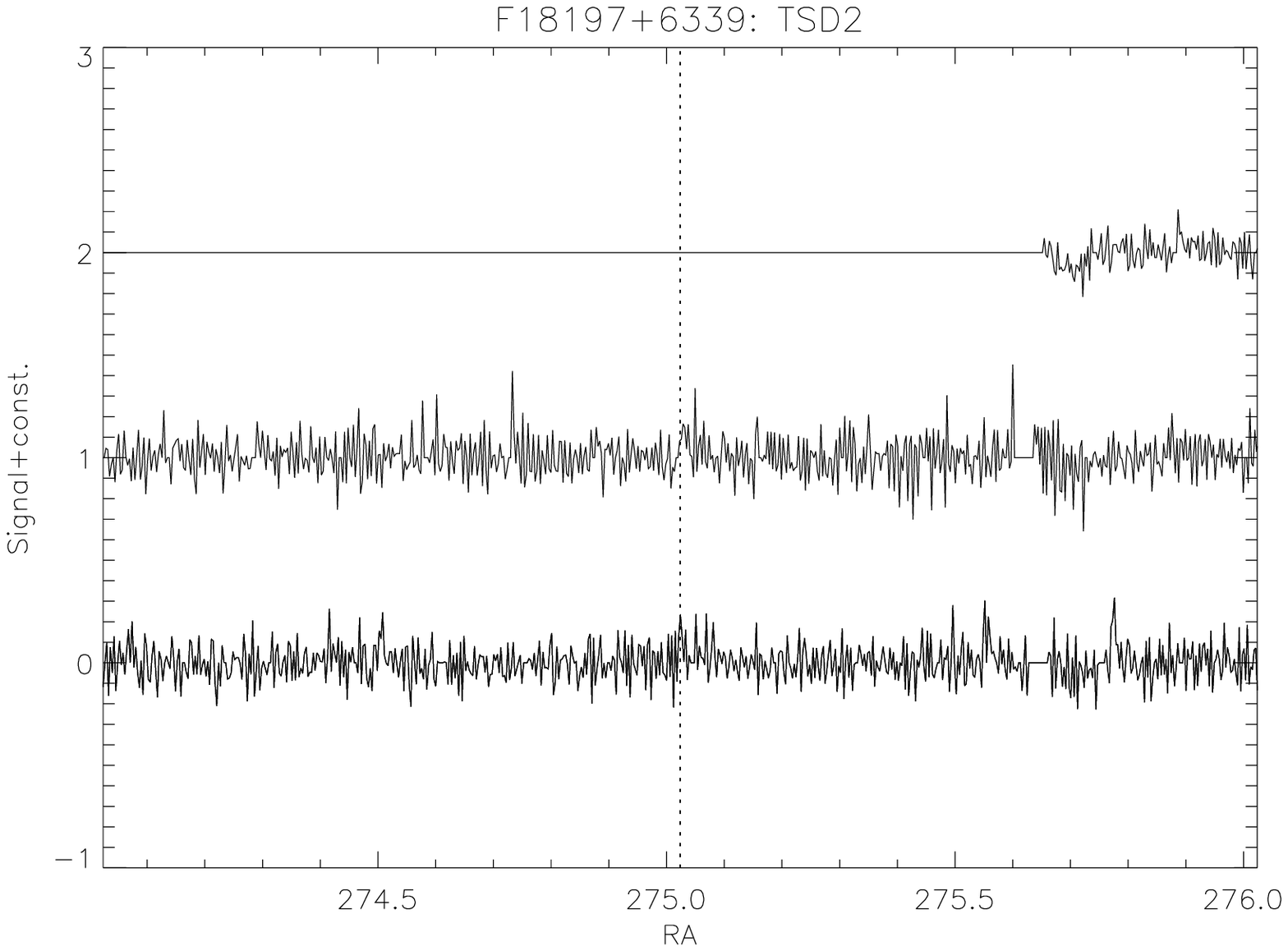}
\includegraphics[height=2.0in,width=3.2in]{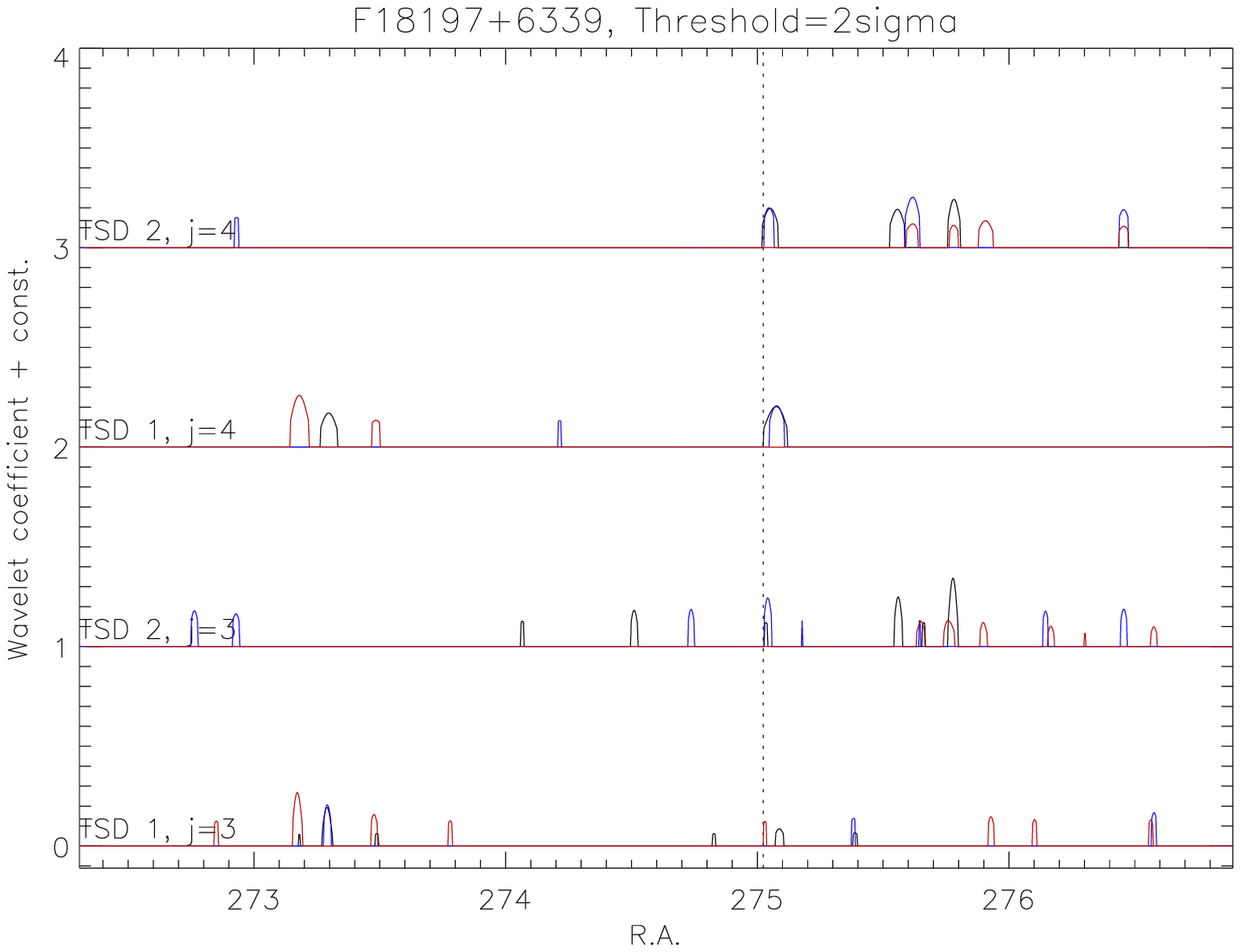}
\includegraphics[height=2.0in,width=3.2in]{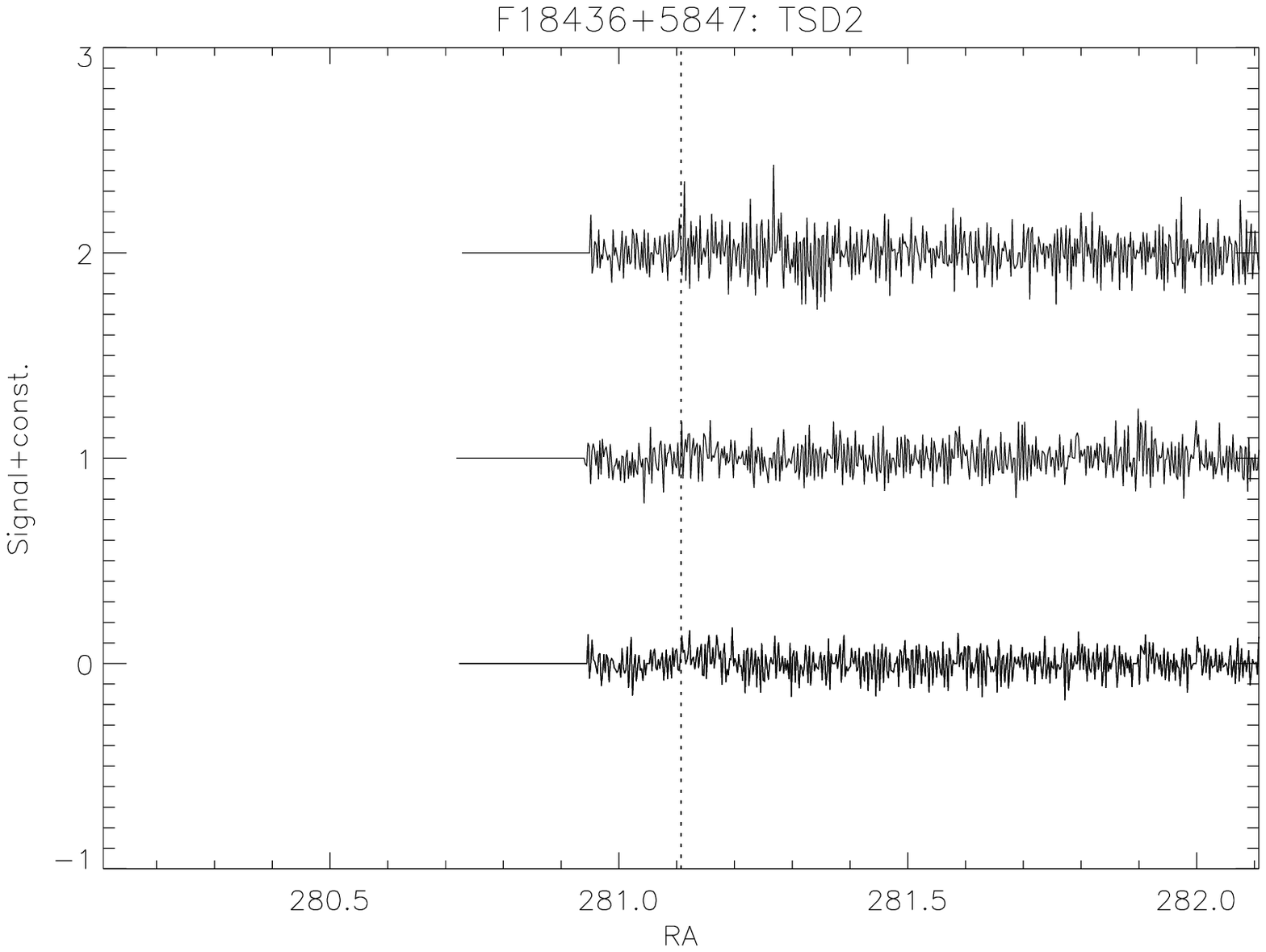}
\includegraphics[height=2.0in,width=3.2in]{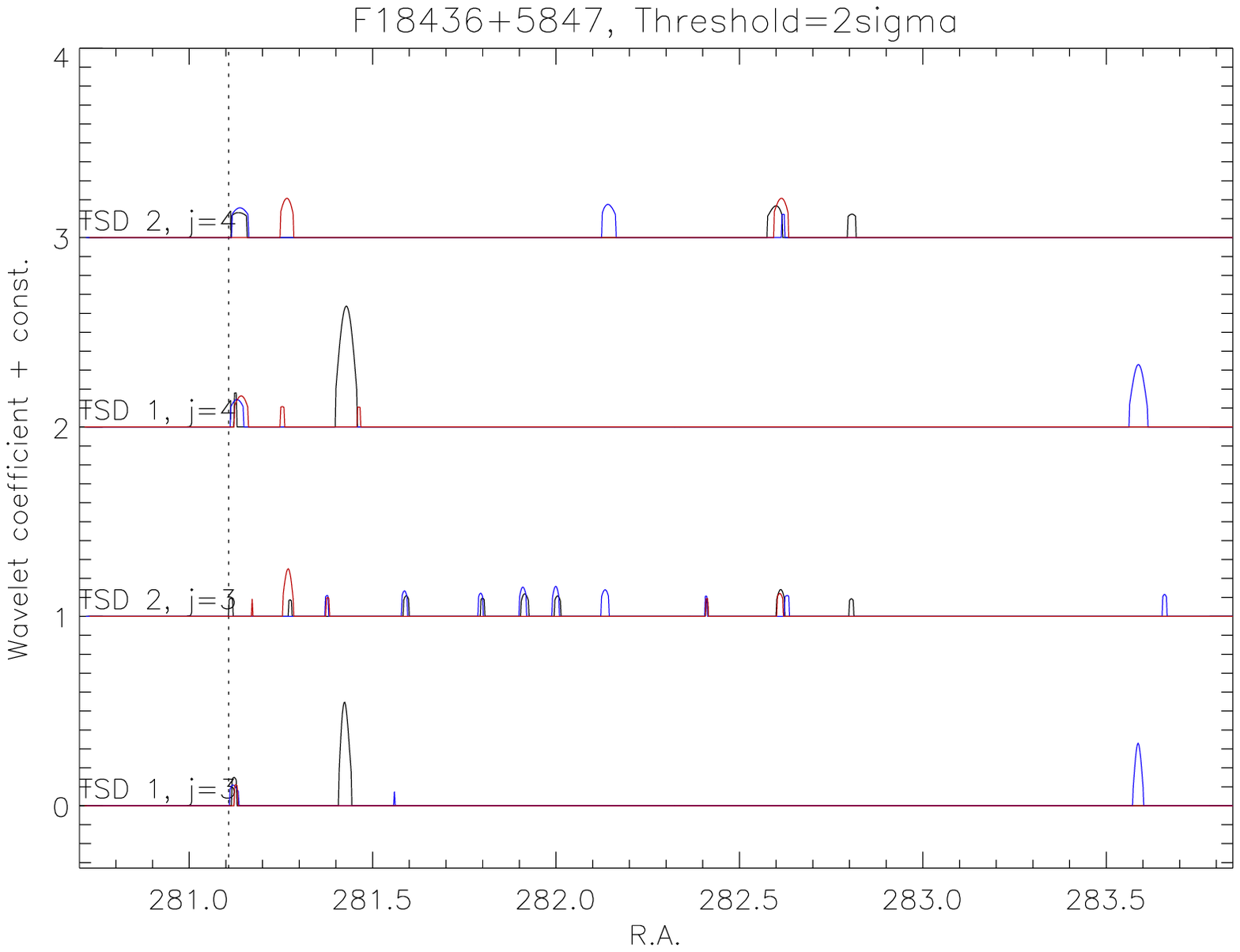}
\includegraphics[height=2.0in,width=3.2in]{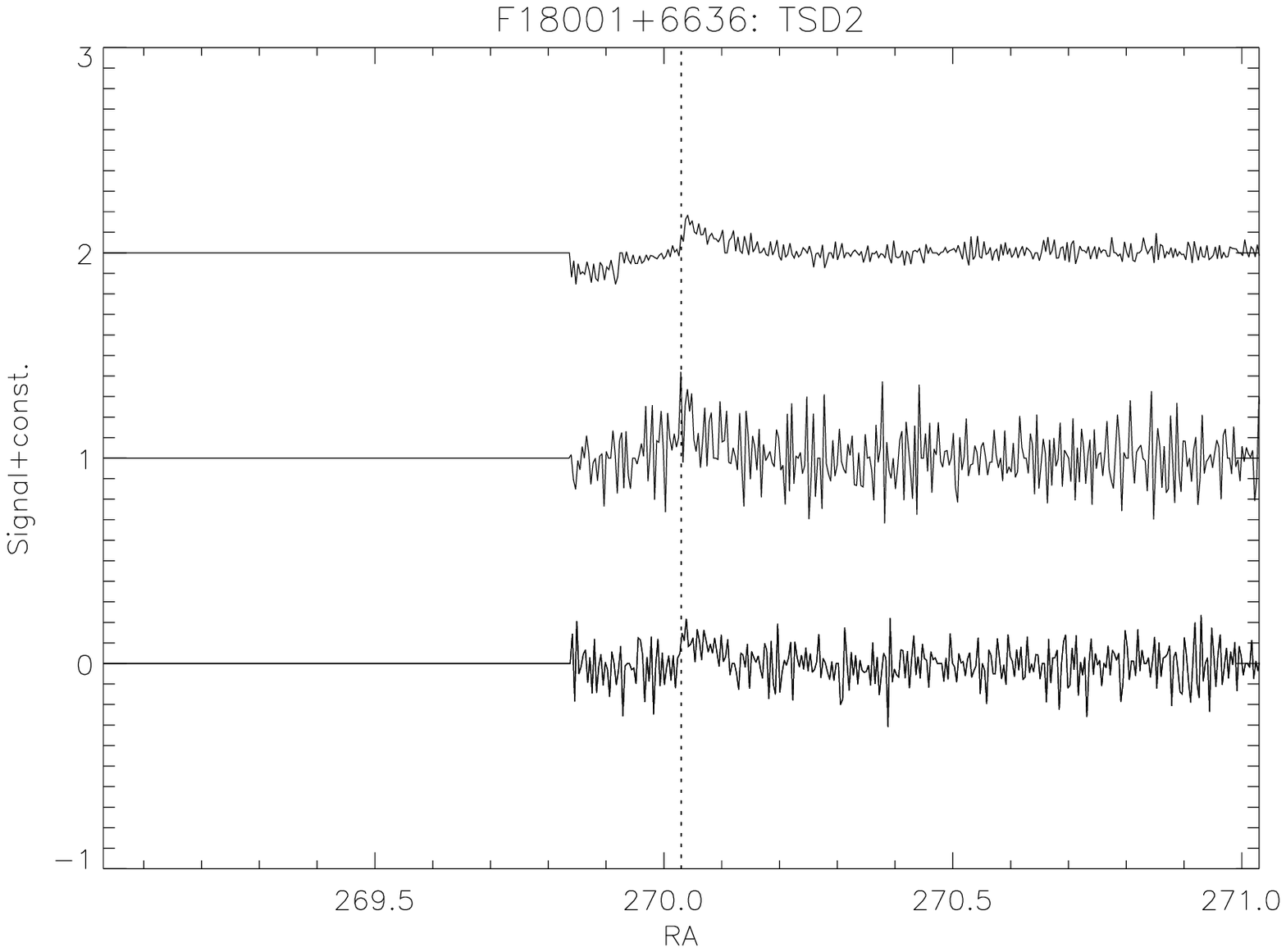}
\includegraphics[height=2.0in,width=3.2in]{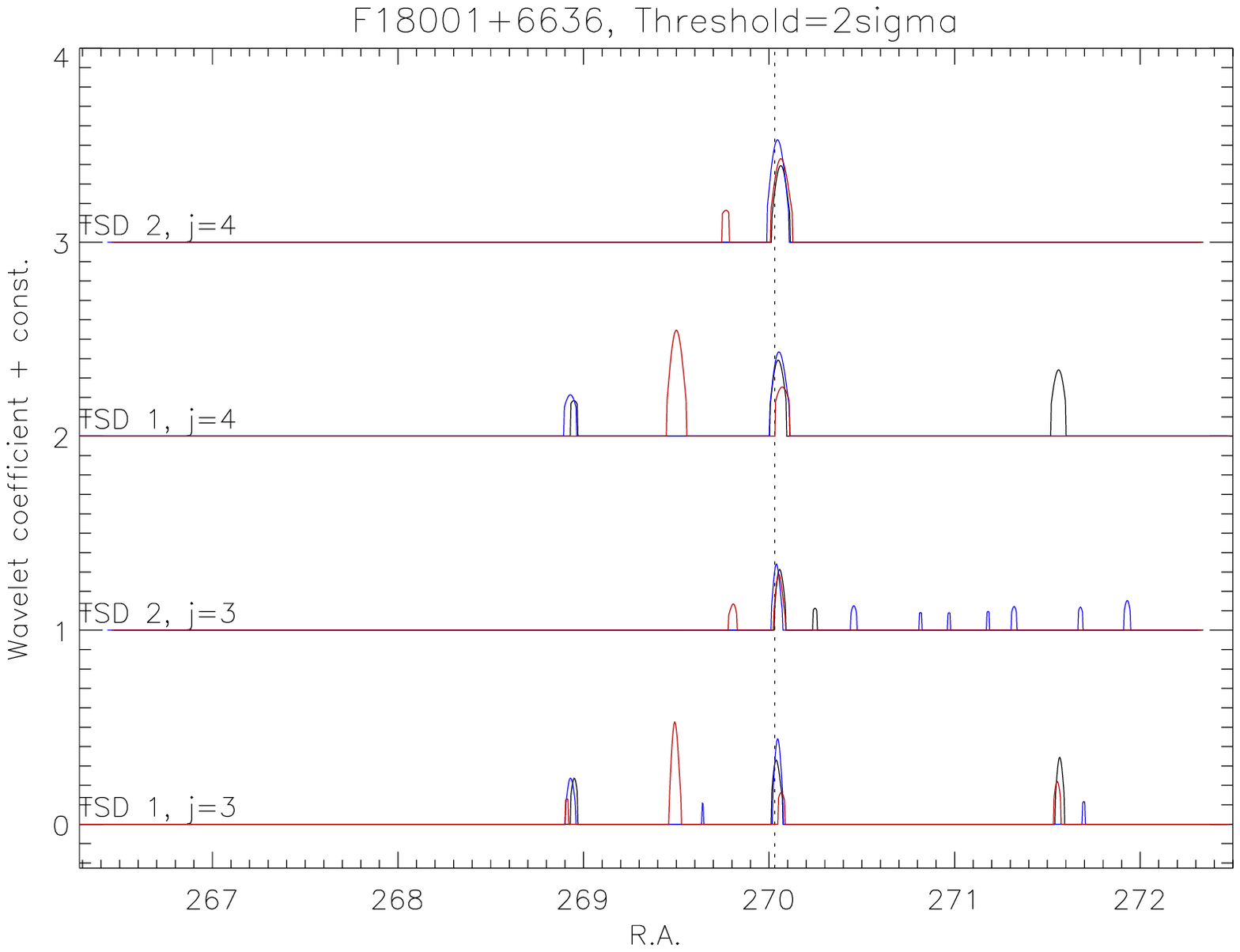}
\caption{Left: examples of the original timelines in the 3 passing detector pixels in one TSD for each IRAS source. Right: the continuous wavelet transform of the entire timeline in each passing detector pixel after thresholding on the Point Source Scales in 2 overlapping TSDs for each source. The dotted line shows the position of the target source.}
\label{fig:orig_timeline}
\end{figure*}

\subsection{Wavelet multiscale representation: definition and outline}
On what scales do we expect point sources to appear? In the WIDE-S 90\,$\mu$m band, the FWHM of the PSF is $\sim$39 arcseconds, the sampling rate is 25.28 Hz, and the scan speed is $\sim$3.6 arcmin per second. Therefore, a single point source can be observed by $\sim$9 samplings in one detector pixel provided no data points are flagged out, 
\begin{equation}
\frac{2 \times 39 ~arcsec}{3.6 \times 60 ~arcsec / sec} \times 25.28 ~Hz \approx 9.
\end{equation}
Therefore, the closest wavelet width to 9 is $2^3=8$ under a dyadic sampling strategy. In other words, the optimal resolution level\footnote{This is the Object Scale defined as the scale where the maximum wavelet coefficient of an object lives in the Multiscale Vision Model (Bijaoui \& Ru\'{e} 1995; Stark \& Murtagh 2002)} for detecting point-like objects is $a=2^{3}$. It follows that quasi-point sources or slightly extended sources mostly inhabit on scale $2^{4}$. To sum up, our prior knowledge of point source detection in the multiscale representation model includes: 

P1. At resolutions $2^{1}$ and $2^{2}$, signals are assumed to be dominated by random noise; 

P2. Point sources can be most easily detected at resolution $2^{3}$. Quasi-point sources / slightly extended sources prefer to show up at resolution $2^{4}$. We will refer to these resolution scales ($2^{3}$ and $2^{4}$) as the Point Source Scales; 

P3. At resolutions coarser than $2^{4}$, cirrus emission and the sky background begin to dominate.   

We use the IRAS sources in Table \ref{IRAS_4sigma} to demonstrate source detection with wavelet multiscale decomposition technique and also to compare with the cirrus candidates. A summary of the procedures is the following.

1. In each detector row\footnote{The FIS has 3 detector rows in the WIDE-S band, each of which has 20 detector pixels (Table 1).},  we search for the detector pixel which passes a target source. The 3 passing detector pixels should be approximately 20 or 21 detectors away from each other due to the configuration of the detector arrays of FIS and the rotation angle ($26.5$ deg) with respect to the scanning direction.  

2. In each detector pixel, the number of data points $N$ is required to be a power of 2, i.e. $N=2^J$. Therefore, we add an adequate amount of zeros at the end of each timeline.  

3. We compute the CWT of the timelines on the Point Source Scales ($2^{3}$ to $2^{4}$), producing a set of wavelet coefficients $w_{a, b}$ on each scale.

4. We calculate the standard deviation $\sigma (a)$ of the wavelet coefficients at each scale and then set any wavelet coefficient to zero if $w_{a, b} < k \sigma (a)$ (insignificant wavelet coefficient). $k$ is chosen to be 2 in this paper. This is known as $hard~thresholding$. 

5. Finally we use seconds- and hours-confirmation to discriminate real sources from spurious ones. This is achieved by comparing the wavelet transform after thresholding in different detector rows (seconds-confirmation) and different TSDs (hours-confirmation). 

\begin{figure*}
\includegraphics[height=4.0in,width=5.7in]{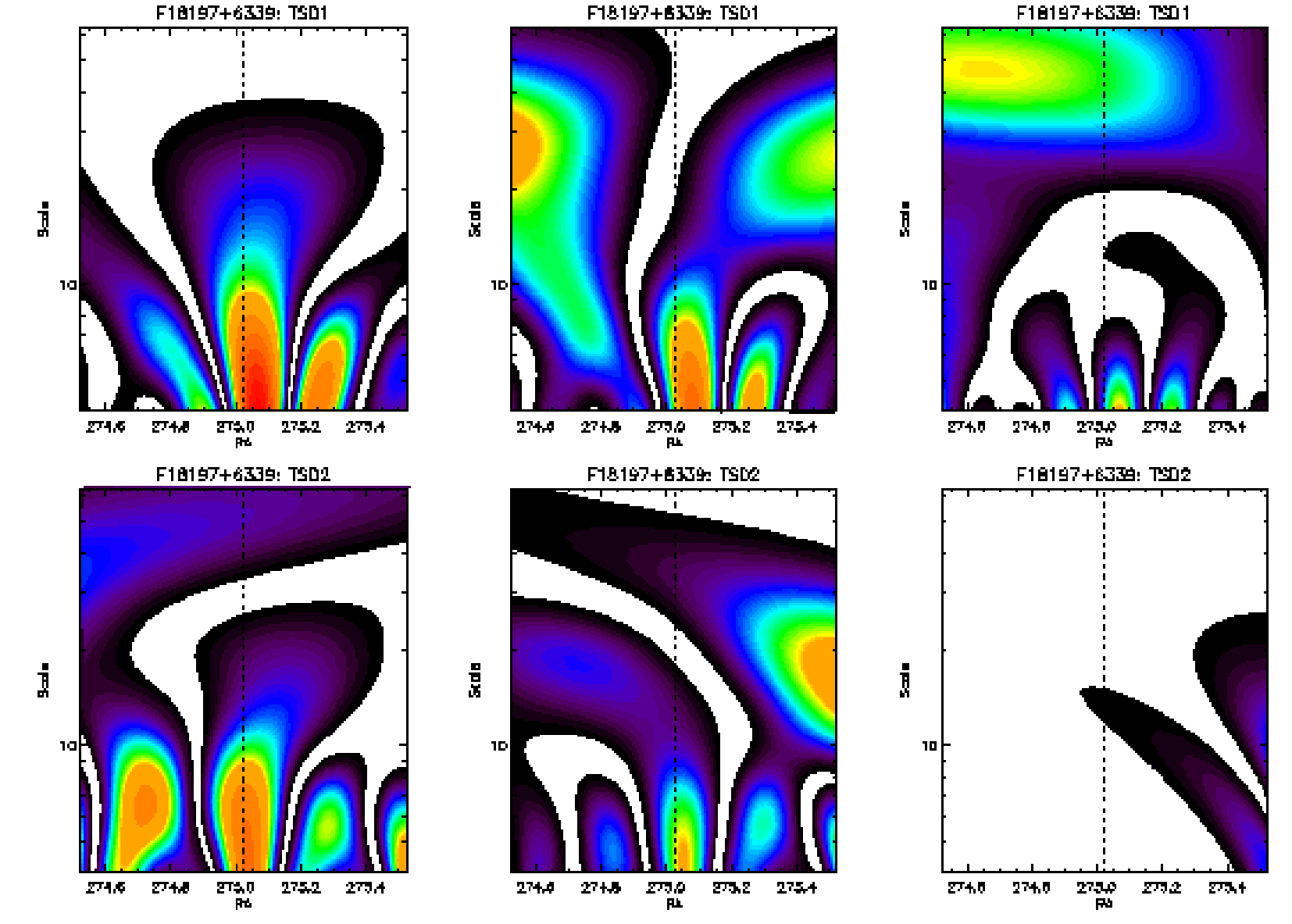}
\caption{The local wavelet power spectrum. The x-axis is centred at the position of the target source (the vertical dotted line).}
\label{fig:contourF1}
\end{figure*}

\subsection{Multiscale analysis of the IRAS sources}
\label{wavelet detection}

Examples of the original timelines in one scan for the 3 IRAS sources in Table 5 are shown in Fig. \ref{fig:orig_timeline}. For each source, there are 3 timelines in the 3 passing detector pixels respectively. The position of the target source is indicated by the intersection between the signal and the vertical dotted line. The brightest source F18001+6636 with$F_{90}^{int}=2.3$ Jy can just be detected by eye, while the other two sources with $F_{90}^{int}=1.0$ Jy are obscured by the noise. 

In the same figure, we have shown the continuous wavelet transform of the timelines after thresholding for each IRAS source in 2 overlapping TSDs (shown as TSD 1 and TSD 2) on scales $2^3$ and $2^4$. In fact, the reconstructed timeline on each scale is proportional to the amplitude of the wavelet coefficient (Torrence \& Compo 1998). The timelines in the 3 passing detector pixels are represented by 3 different colors, black, red and blue. If more than two positive signals of different colours overlap at the same position in one TSD, then it is regarded as a seconds-confirmed detection. To discriminate real sources from spurious ones, we also require hours-confirmation by checking the other TSD at the same position. Only when a seconds-confirmed detection is present in both TSDs, i.e. a seconds- and hours-confirmed source, do we acknowledge it as a real source. 

So, the first conclusion from Fig. \ref{fig:orig_timeline} is that each IRAS source can be clearly identified as the only seconds- and hours-confirmed detection in the entire TSD at resolution $2^{4}$. Secondly, although F18001+6636 is also present with seconds- and hours-confirmation at resolution $2^{3}$, the two fainter sources F18197+6339 and F18436+5847 are not detected according to our selection criteria at $j=3$. In addition, due to the noisy nature of the data, F18197+6339 and F18436+5847 will be missing at $j=4$ if a higher threshold is used.  

Fig. \ref{fig:contourF1} shows the WPS as a function of position and scale for IRAS source F18197+6339. Again, there are 6 panels (3 passing detector pixels per TSD $\times$ 2 TSDs) and the vertical dotted line in the middle of each panel marks the position of the target source. In order to compare the WPS in different detector pixels and for different objects, we use common contour levels in Fig. \ref{fig:contourF1}, Fig. \ref{fig:contour_C1C2}, Fig. \ref{fig:contour_C3C4} and Fig. \ref{fig:contour_C5C6}. The wavelet power increases towards red in each contour plot. A high power at the position of the IRAS source and on the correct scales (around $2^3=8$) can be unmistakably identified in almost every contour panel of F18197+6338. The missing power in the rightmost panel in TSD 2 is due to the flagged data around the source (see the top left panel in Fig. \ref{fig:orig_timeline}). The noisy nature of the data can been seen in the contour plots of the wavelet power. For example, there are contours of large power in the timelines on a range of scales including the Point Source Scales.

\begin{figure*}
\includegraphics[height=2.2in,width=2.2in]{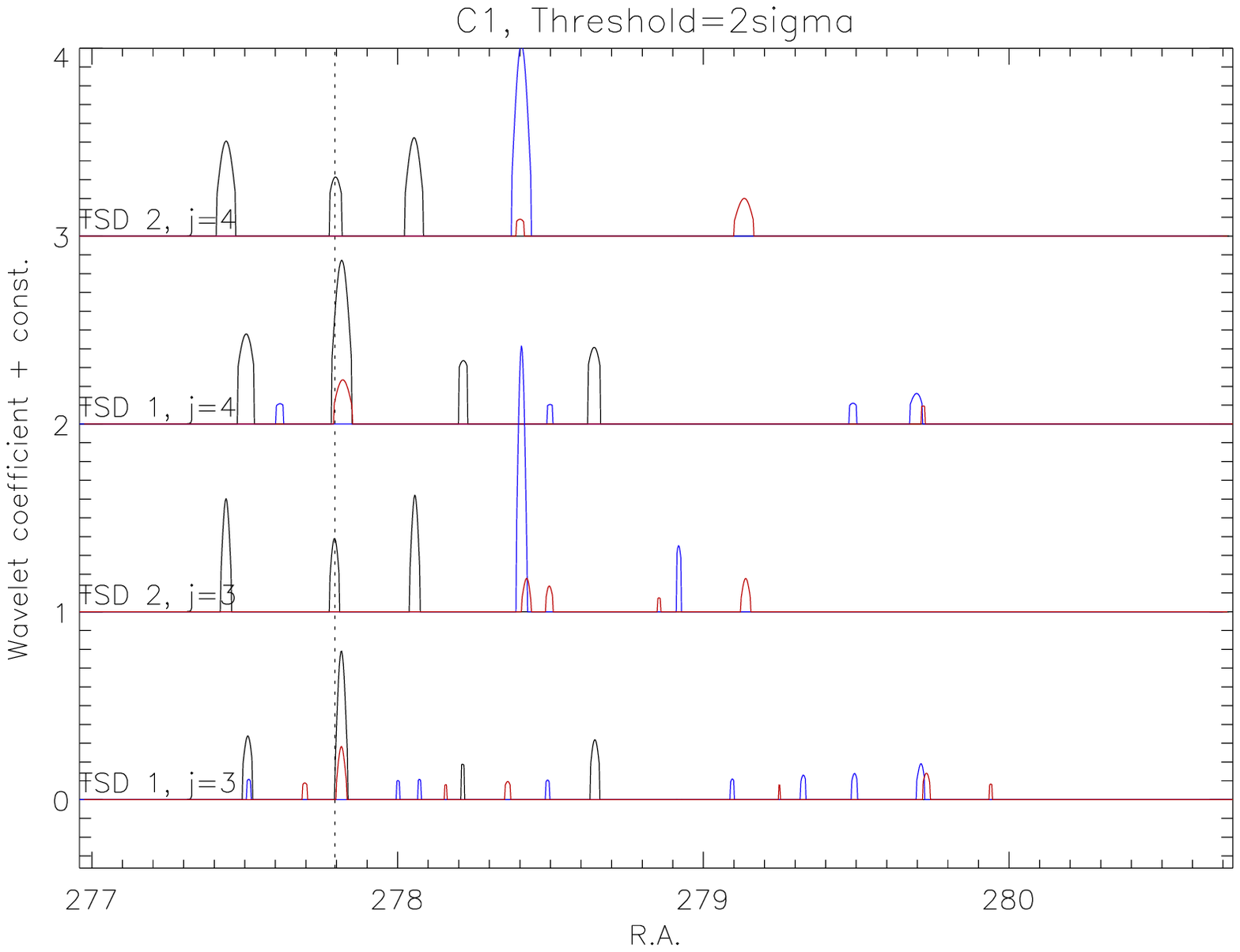}
\includegraphics[height=2.2in,width=2.2in]{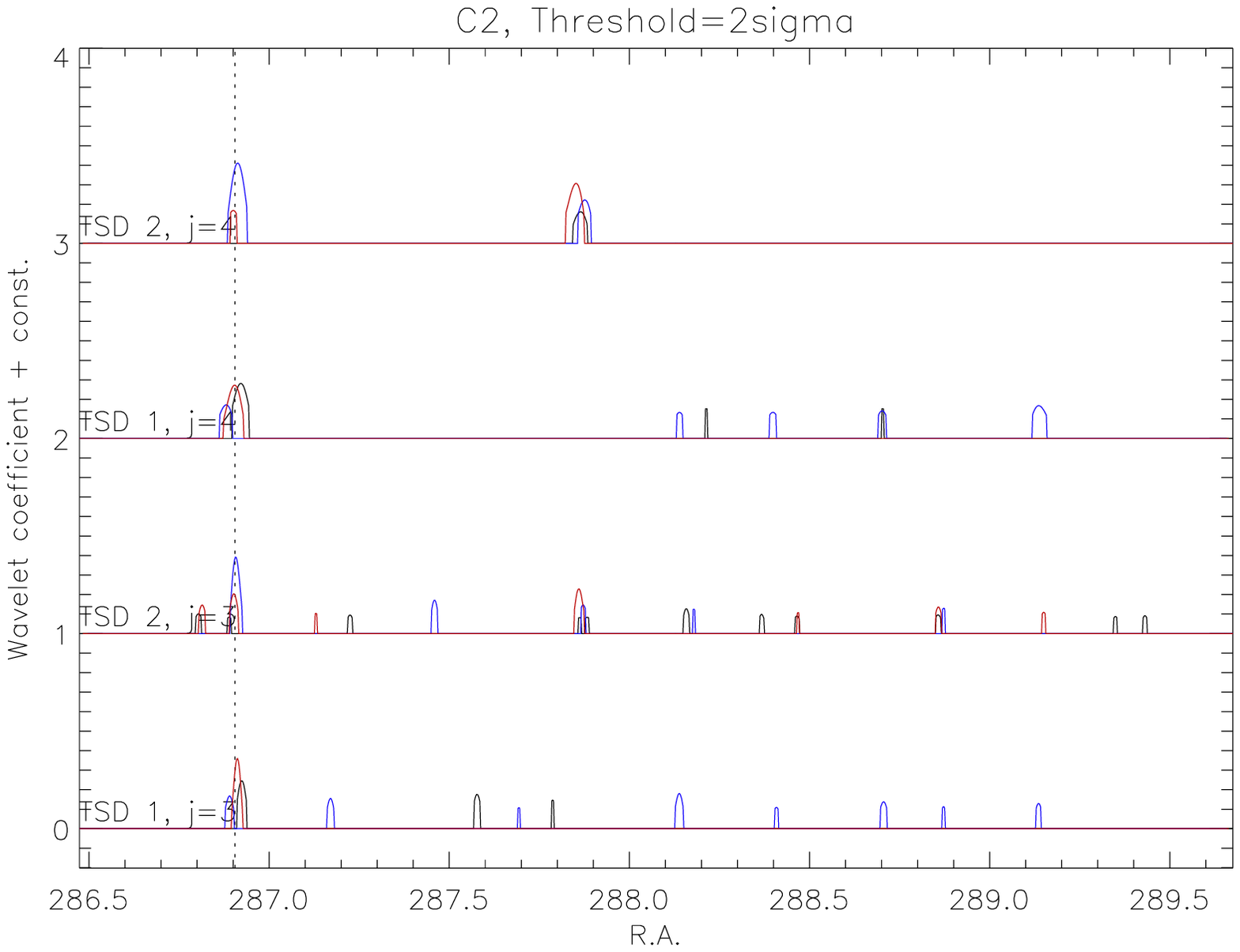}
\includegraphics[height=2.2in,width=2.2in]{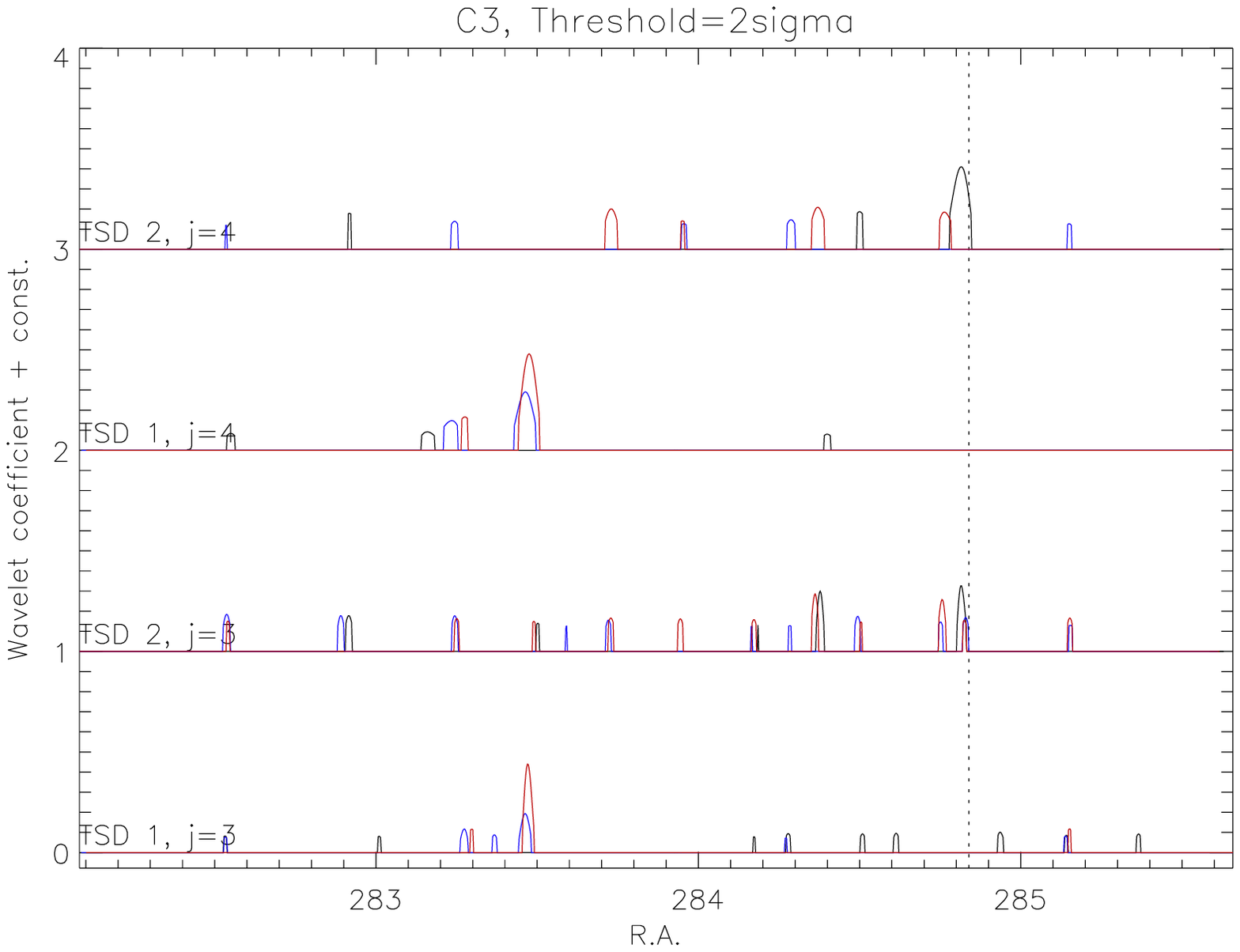}
\includegraphics[height=2.2in,width=2.2in]{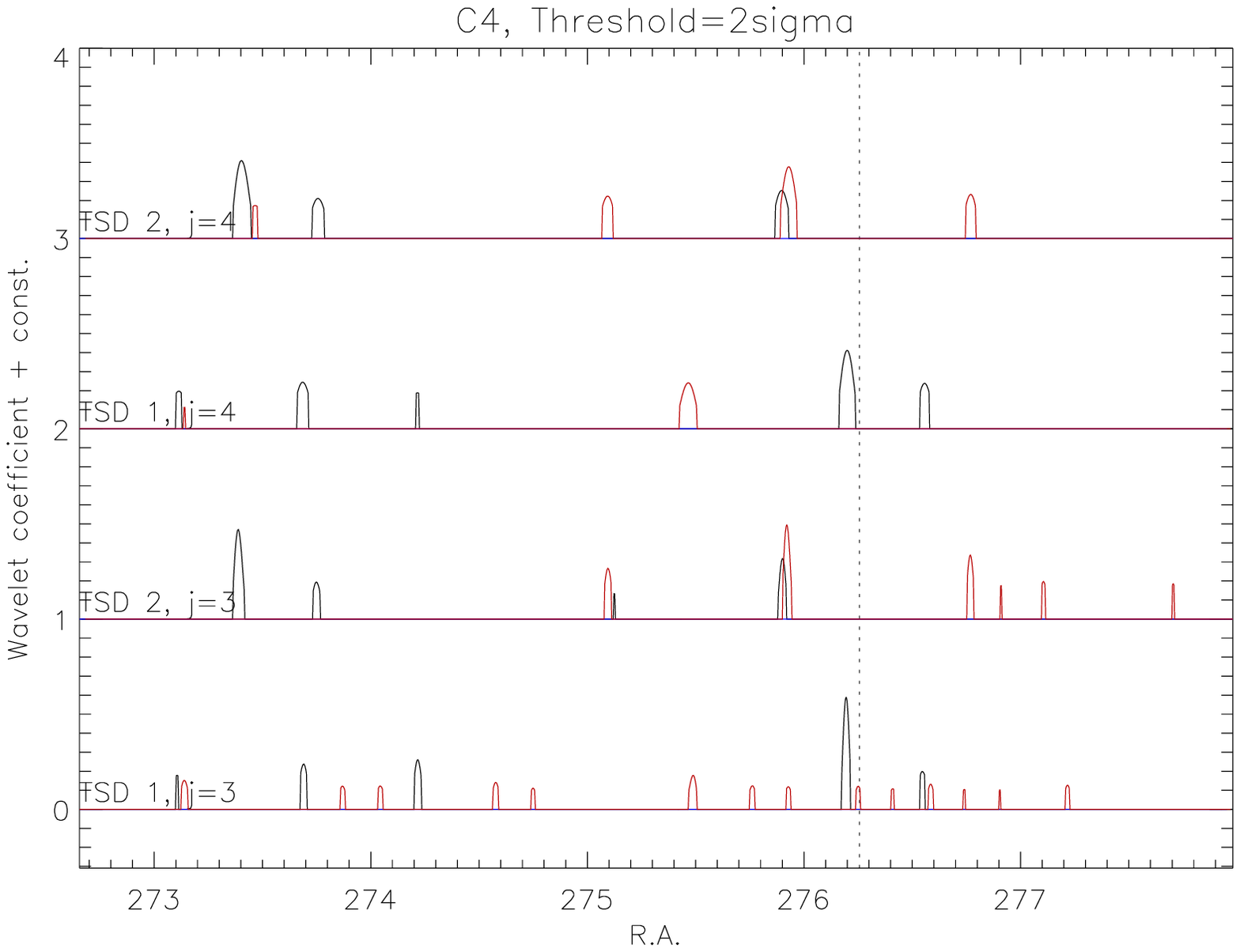}
\includegraphics[height=2.2in,width=2.2in]{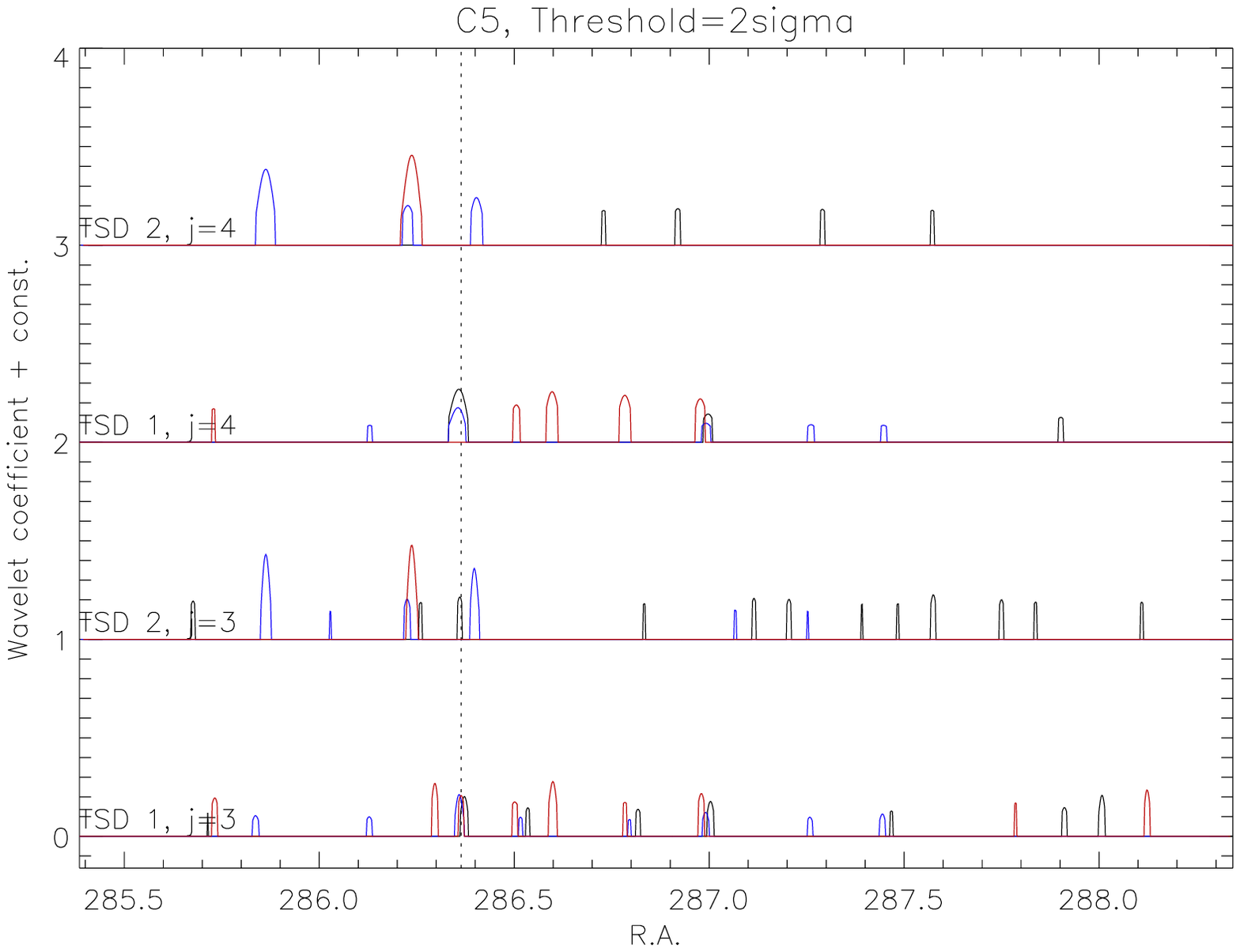}
\includegraphics[height=2.2in,width=2.2in]{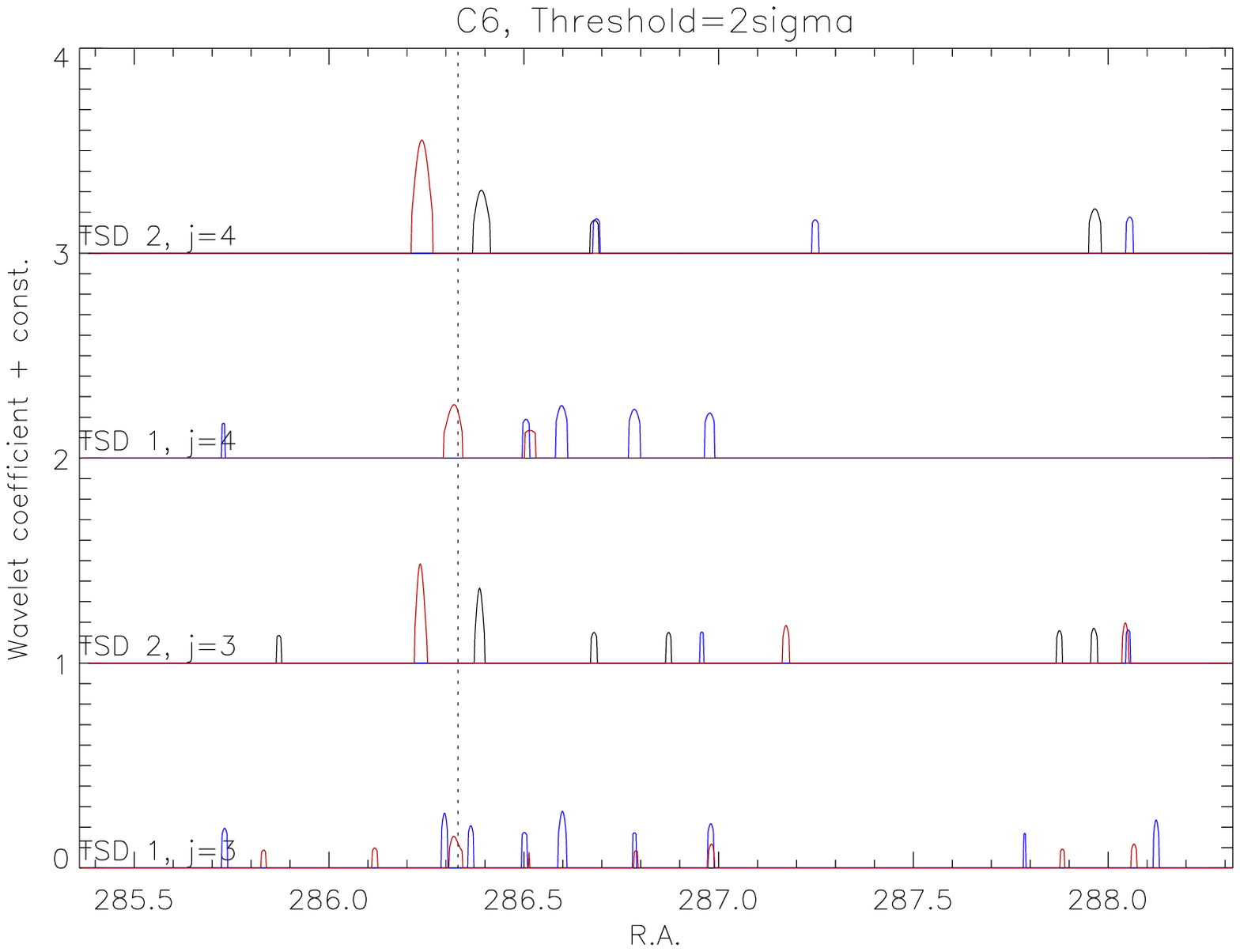}
\caption{The continuous wavelet transform after thresholding for cirrus candidates C1 to C6. Each panels shows the wavelet coefficients in 3 passing detector pixels in 2 overlapping TSDs. The position of the cirrus candidate is indicated by the dotted line.}
\label{fig:CWT_cirrus}
\end{figure*}

\subsection{Multiscale analysis of the cirrus candidates}
\label{wavelet_cirrus}

The same wavelet transform procedure has been applied to the 6 cirrus candidates selected with S/N$>4$ and the result is shown in Fig.~\ref{fig:CWT_cirrus}. Apart from candidate C2, all the other cirrus candidates are clearly not seconds- and hours-confirmed detections on the Point Source Scales. Therefore, according to our selection criteria these sources are spurious and not new discoveries by AKARI. 

\begin{figure*}
\includegraphics[height=2.2in,width=3.4in]{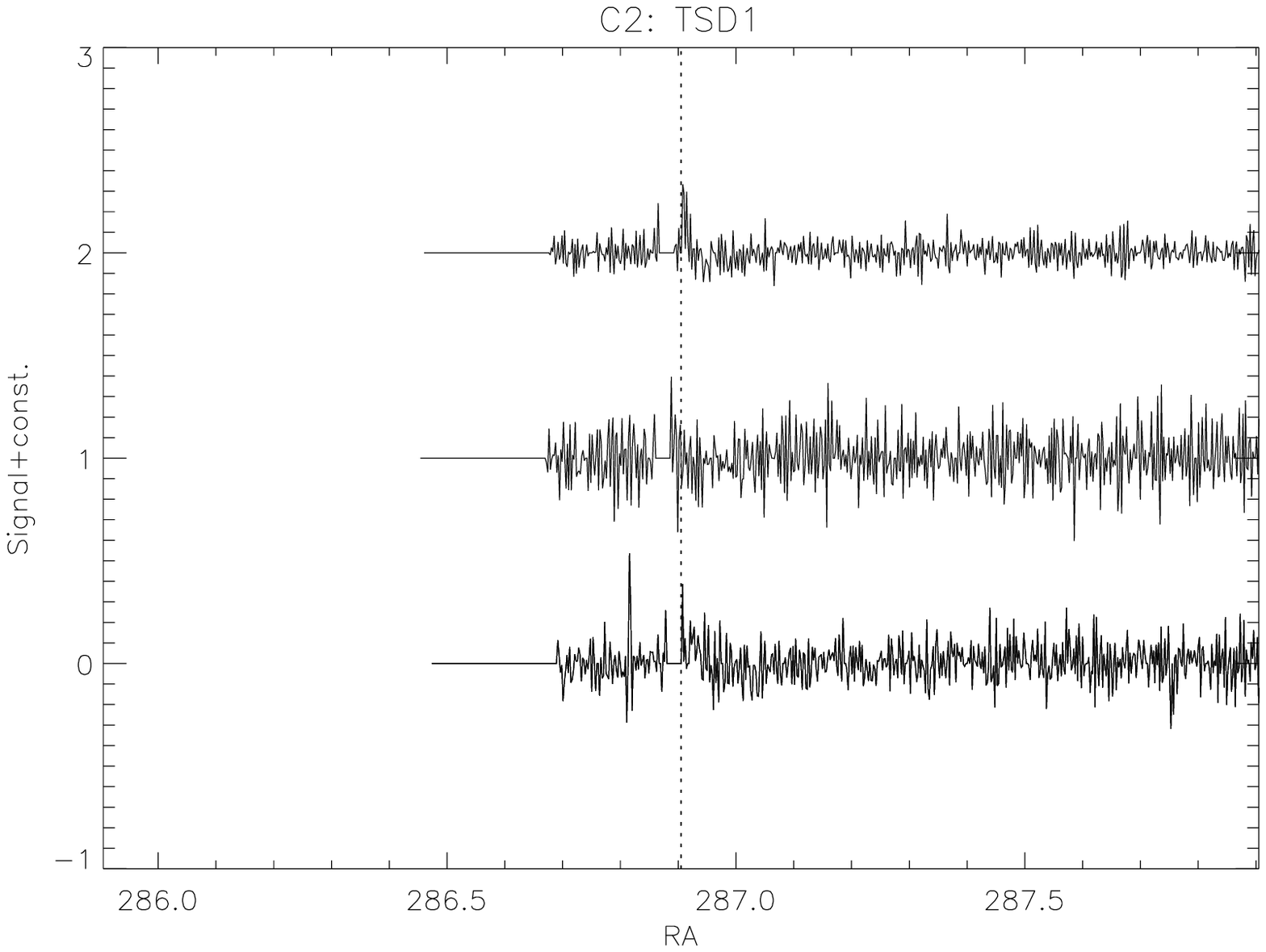}
\includegraphics[height=2.2in,width=3.4in]{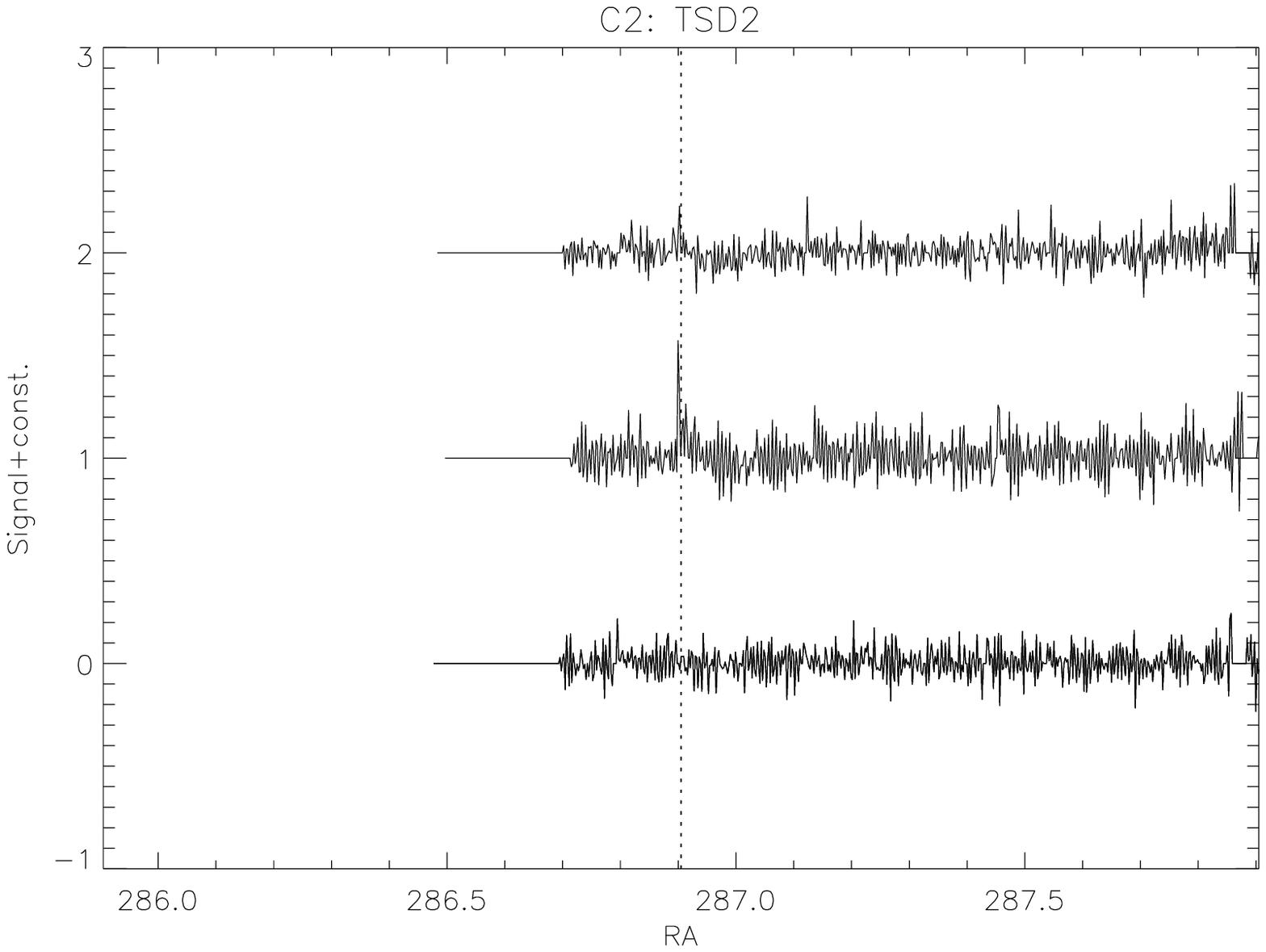}
\caption{The original timelines of cirrus candidate C2 in the 3 passing detector pixels in 2 overlapping TSDs.}
\label{fig:orig_C2}
\end{figure*}

\begin{figure*}
\includegraphics[height=4.0in,width=6.0in]{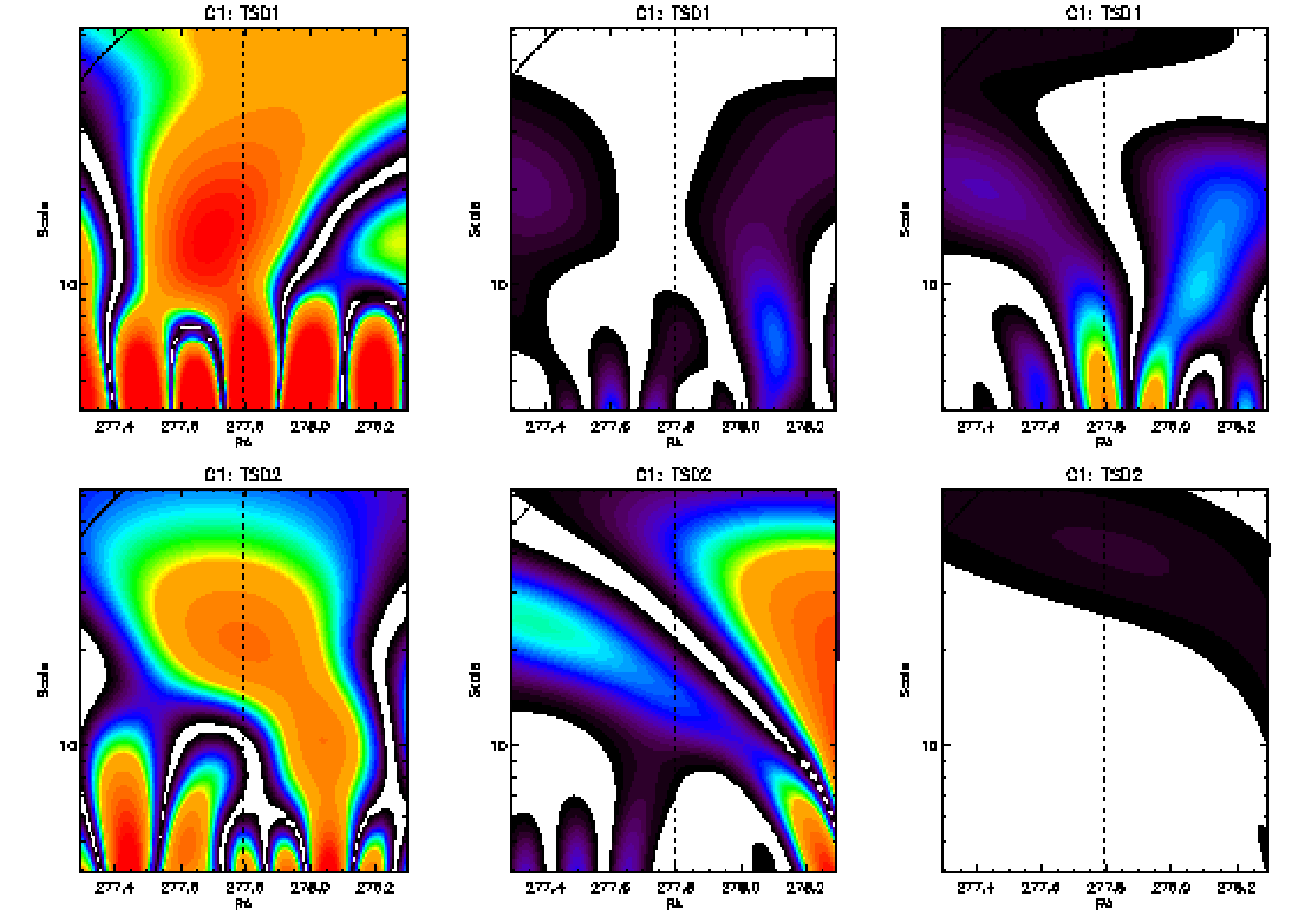}
\includegraphics[height=4.0in,width=6.0in]{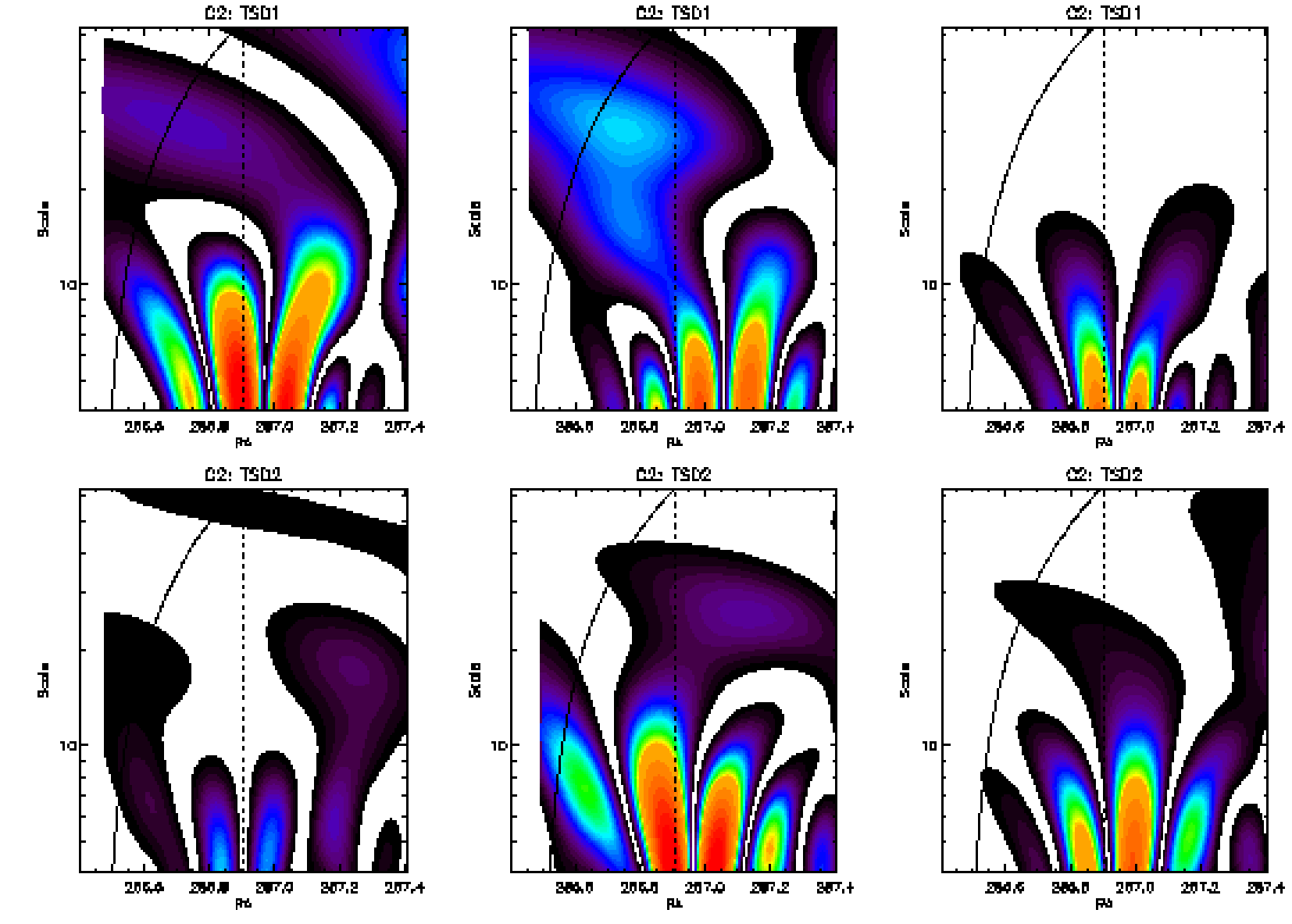}
\caption{The local wavelet power spectrum of cirrus candidate C1 and C2 in 2 TSDs from scale $2^2$ to $2^6$. The x-axis is centred at the position of the target source (dotted line).}
\label{fig:contour_C1C2}
\end{figure*}

In Fig.~\ref{fig:orig_C2}, the original timelines of cirrus candidate C2 in 2 TSDs (TSD 1 and TSD 2) are shown. The bottom and top timeline in TSD 1 and the middle timeline in TSD 2 clearly show the presence of a spike at the position of the cirrus candidate. The contour plots of the WPS of C2 are shown in the lower half of Fig.~\ref{fig:contour_C1C2}. In both TSDs, the effect of a spike is manifested by contours of power resembling that of point sources at the vertical dotted line (the left and right panel in C2 TSD 1 and the middle panel in C2 TSD 2). For other cirrus candidates, contour plot similar to that of a point source can be seen in some detector pixels. An investigation of the original timelines shows that it is also due to spikes found at the position of the cirrus candidates. 

\begin{figure*}
\includegraphics[height=4.0in,width=6.0in]{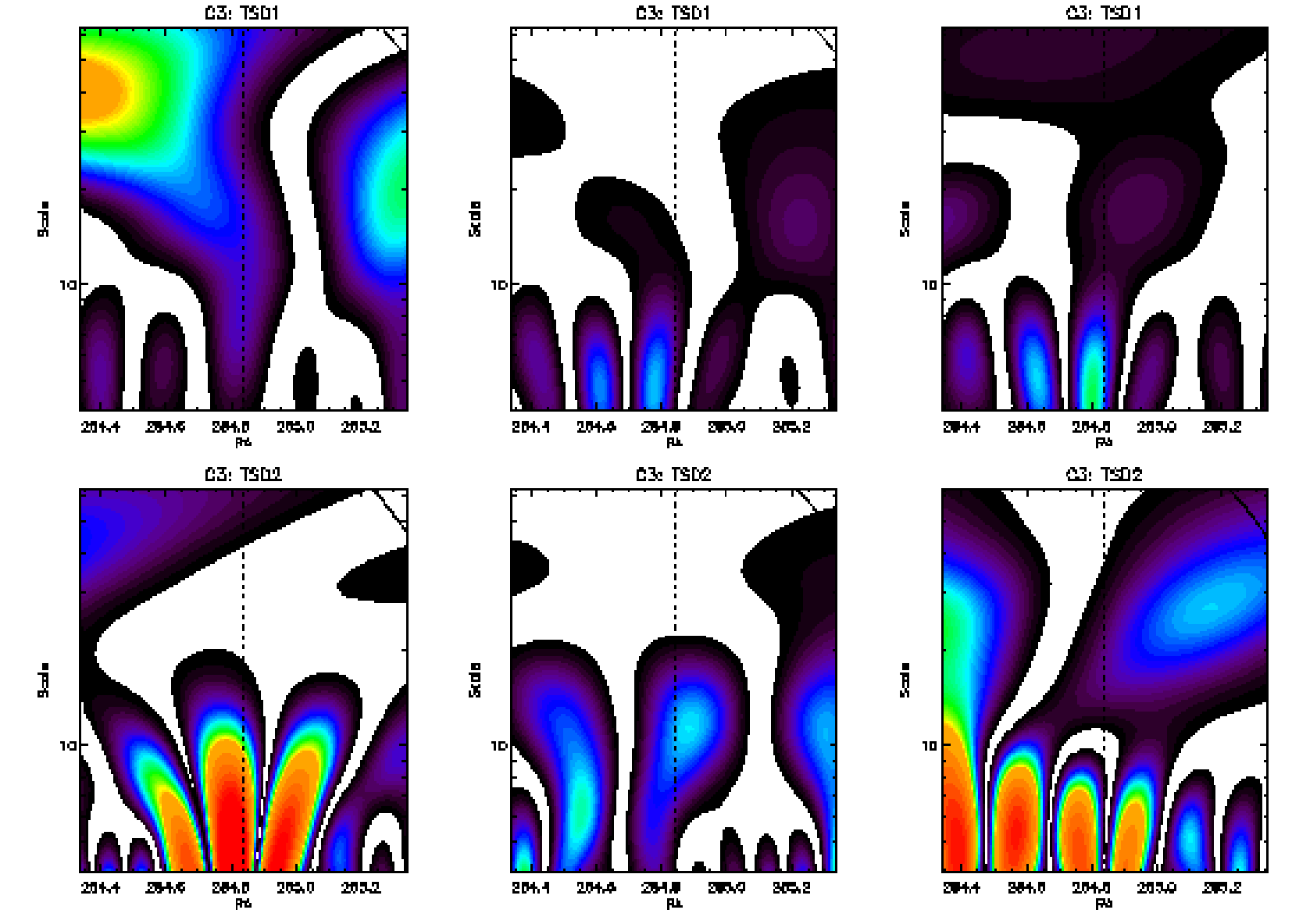}
\includegraphics[height=4.0in,width=6.0in]{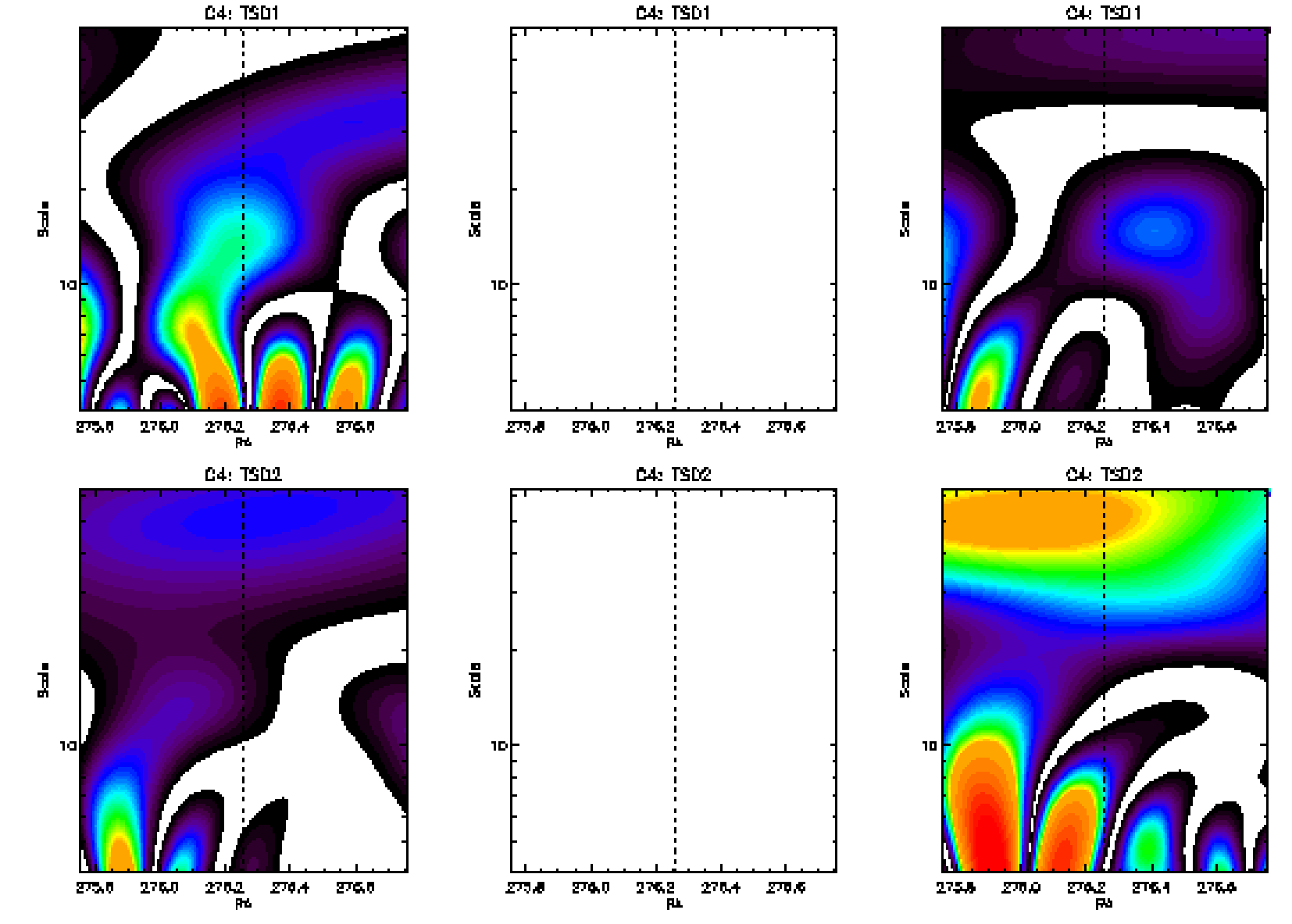}
\caption{The local wavelet power spectrum of cirrus candidate C3 and C4 in 2 TSDs from scale $2^2$ to $2^6$. The x-axis is centred at the position of the target source (dotted line).}
\label{fig:contour_C3C4}
\end{figure*}

\begin{figure*}
\includegraphics[height=4.0in,width=6.0in]{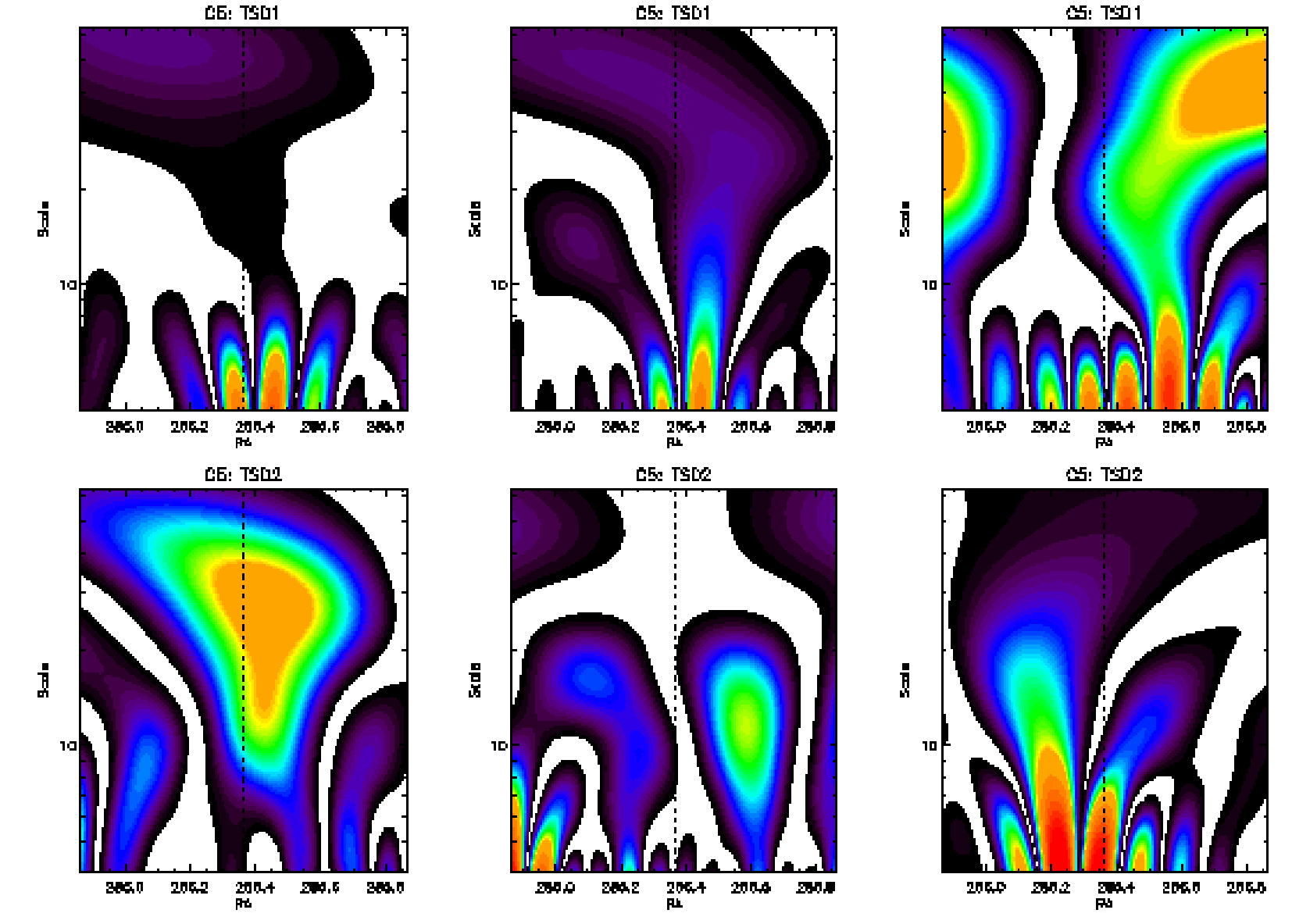}
\includegraphics[height=4.0in,width=6.0in]{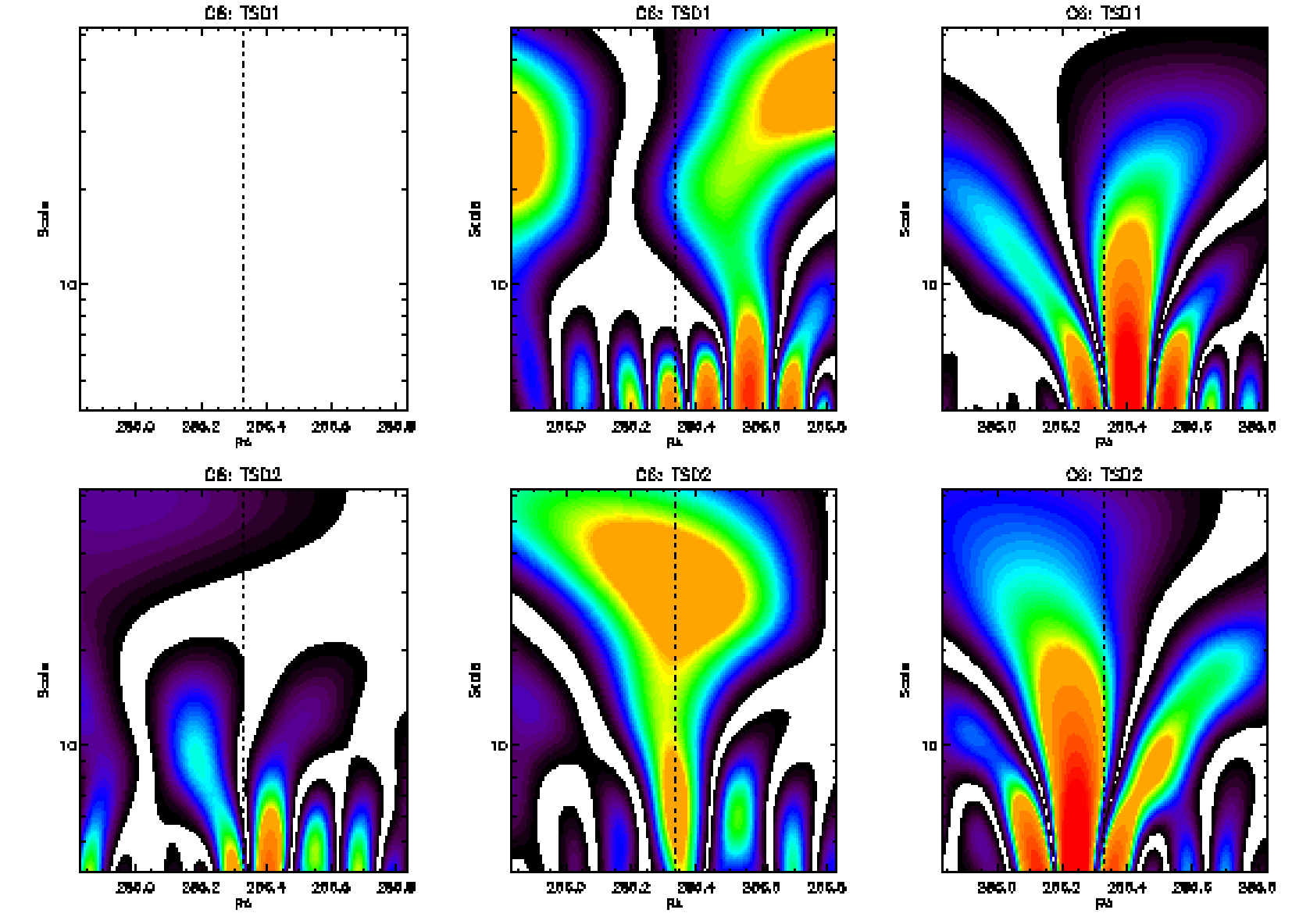}
\caption{The local wavelet power spectrum of cirrus candidate C5 and C6 in 2 TSDs from scale $2^2$ to $2^6$. The x-axis is centred at the position of the target source (dotted line).}
\label{fig:contour_C5C6}
\end{figure*}

If these non-IRAS detections detected in Timeline Analysis with seconds- and hours-confirmation are induced by extended structures such as cirrus clouds, they should appear as confirmed detections on scales larger than the Point Source Scales. However, for every cirrus candidate studied in this paper, this assumption can be rejected based on the WPS plots from Fig. \ref{fig:contour_C1C2} to Fig. \ref{fig:contour_C5C6}. In some contour plots a thin curved line can be seen at either side of the panel (C1, C2 and C3). Regions above the thin line are the cone of influence (Torrence \& Compo) where the edge effect of the TSD makes the power spectrum unreliable. In the case of cirrus candidate C1, regions of high power on large scales are present in the left panel of TSD 1 as well as TSD 2 at the vertical dotted line, while they are absent in other detector pixels. Similarly, for cirrus candidates C4, C5 and C6, at the position of the candidate, only one detector pixel display high power on large scales in each case (the right panel in C4 TSD 2; the left panel in C5 TSD 2; the middle panel in C6 TSD 2). For C2 and C3, apart from spikes seen in some detector pixels, the dotted linen in does not intersect with regions of large power. The fact that the contour plots of wavelet power along the dotted line look wildly different across different detector pixels of the same object leads us to conclude that these spurious sources are not caused by cirrus emission but random noise in the timelines.

\section{DISCUSSIONS AND CONCLUSIONS}
\label{conclusions}
The AKARI All-Sky Survey has covered over $90$ per cent of the entire sky twice. So, theoretically the $5\sigma$ point source flux detection limit at the WIDE-S 90\,$\mu$m band will reach $1.1$ Jy$~/~\sqrt{2~(overlapping~scans)~\times~3~(detector~rows)}\approx0.4$ Jy assuming equal data quality among different detector rows and different scans, which will give rise to more than 40,000 extragalactic sources over the whole sky. In regions covered by multiple scans (e.g. the NEP), the point source detection limit will be improved further. Therefore, future catalogues from the AKARI All-Sky Survey will be valuable samples for large-scale structure studies. Further improvement on the absolute calibration can be made with better quality data from the data reduction pipeline which is under intensive development, larger data sets and well-modelled objects (comets, stars, galaxies etc.). 

The localising property of wavelet transform prevents transient signals being submerged in the average characteristics of the overall timeline. Wavelet transform, as a band-pass filter, decomposes data onto a succession of resolution levels and hence separates signal components due to physical structures of different sizes. In our wavelet multiscale analysis, IRAS sources are the only detections with seconds- and hours-confirmation at the expected resolutions for point sources or small extended sources. However, as our data set only provides us with 3 sources above a $4\sigma$ threshold, a larger sample will be desirable to further test our method. 

By thresholding at a high signal-to-noise ratio (S/N $>4$ or $5$), we find that most of the seconds- and hours-confirmed sources which are not in the IRAS point source catalogues trace cirrus clouds, i.e. they reside in environments with strong or intermediate 100\,$\mu$m emission. In the co-added AKARI 90\,$\mu$m maps, it is clearly shown that these non-IRAS sources referred to as cirrus candidates are associated with bright extended structures, either in the in-scan direction (the time direction) or both the in-scan and cross-scan direction. Unlike the IRAS sources, most of the cirrus candidates fail to to be detected with seconds- and hours-confirmation on the Point Source Scales under continuous wavelet transform. It confirms that they have a different scale preference compared with the IRAS sources. However, these non-IRAS point sources are not likely due to cirrus judging from the contour plots of the wavelet power. At the position of each cirrus candidate, usually one or two detector pixels exhibit high power on large scales, while other detectors either show presence of spikes on the Point Source Scales or no regions of high power at all. In other words, the WPS differ significantly across different detector pixels in the same scan and also different scans of the same candidate. With improvement of data processing currently in progress, we can filter out excessive noise in the raw timelines to increase the reliability and completeness of the AKARI All-Sky Survey data products.

\section*{Acknowledgments}
This research is based on observations with AKARI which is managed and operated by the Institute of Space and Astronautical Science (ISAS), Japan Aerospace Exploration Agency (JAXA), with collaboration from universities and research institutes in Japan, the European Space Agency (ESA), the IOSG Consortium with includes Imperial College, UK, Open University, UK, University of Sussex, UK, and University of Groningen, Netherlands, and Seoul National University, Korea. The FIS instrument is developed by Nagoya University, ISAS/JAXA, the University of Tokyo, and the National Astronomical Observatory of Japan and other institutes, with contributions of NICT to the development of the detectors. The UK participation to the AKARI project is supported in part by PPARC/STFC.

This research has made use of the wavelet software provided by C. Torrence and G. Compo, and is available at URL: http://paos.colorado.edu/research/wavelet.

This research has made use of the NASA/IPAC EXTRAGALACTIC DATABASE (NED) which is operated by the 
JET PROPULSION LABORATORY, CALTECH, under contract with the NATIONAL AERONAUTICS AND SPACE ADMINISTRATION.


\begin{thebibliography}{99}
\bibitem{} Bijaoui A., Ru\'{e} F., 1995, Signal Processing, 46, 345-362
\bibitem{} Burrus C.S., Gopinath R.A., Guo H., 1998, Introduction to wavelets and wavelet transforms, Prentice-Hall
\bibitem{} Cay\'{o}n L. et al., 2000, MNRAS, 315, 757
\bibitem{} Efstathiou A. et al., 2000, MNRAS, 319, 1169
\bibitem{} Gautier T.N. III, Boulanger F., P\'{e}rault M., Puget J.L., 1992, AJ, 103, 1313
\bibitem{} Gonz\'{a}lez-Nuevo J., Arg\"{u}eso F., L\'{o}pez-Caniego M., Toffolatti L., Sanz J.L., Vielva P., Herranz D., 2006, MNRAS, 369, 1603
\bibitem{} Helou G., Beichman C.A., 1990, Proc. of the 29th Liege International Astrophysical Coll. ESA Publications Division, Noordwijk, p.117
\bibitem{} Heraudeau Ph. et al., 2004, MNRAS, 354, 924
\bibitem{} Jeong W.-S., Lee H.M., Pak S., Nakagawa T., Kwon S.M., Pearson C.P., White G.J., 2005, MNRAS, 357, 535
\bibitem{} Jeong W.-S., Pearson C.P., Lee H.M., Pak S., Nakagawa T., 2006, MNRAS, 369, 281
\bibitem{} Jeong W.-S. et al., 2007, PASJ, 59, S429
\bibitem{} Kawada M. et al., 2007, PASJ, 59, S389
\bibitem{} Low F.J. et al., 1984, ApJ, 278, L19
\bibitem{} Moshir M. et al., 1992, Explanatory Supplement to the IRAS Faint Source Survey, Version 2, JPL D-10015 8/92 (Pasadena:JPL)
\bibitem{} M\"{u}eller T.G., Lagerros J.S.V., 2002, A\&A 381, 324
\bibitem{} M\"{u}eller T.G. et al., 2008 (in prep.)
\bibitem{} Murakami H. et al., 2007, PASJ, 59, S369
\bibitem{} Onaka T. et al., 2007, PASJ, 59, S401
\bibitem{} Starck J., Murtagh F., 2002, Astronomical Image and Data Analysis, Springer-Verlag Berlin Heidelberg
\bibitem{} Savage R., Oliver S., 2007, ApJ, 661, 1339
\bibitem{} Schlegel D.J., Finkbeiner D.P., Davis M., 1998, ApJ, 500, 525
\bibitem{} Tenorio L., Jaffe A.H., Hanany S., Lineweaver C.H., 1999, MNRAS, 310, 823
\bibitem{} Torrence C., Compo G.~P., 1998, A Practical Guide to Wavelet Analysis. Bull. Amer. Meteor. Soc., 79, 61–78.
\bibitem{} Edited by van der Berg J.~C., 2004, Wavelets in Physics, Cambridge Unversity Press
\bibitem{} Yamamura I. et al., in prep.
\end{thebibliography}
\end{document}